\documentclass[twocolumn, amsmath,amssymb,floatfix]{revtex4}
\usepackage{graphicx}
\usepackage{verbatim}

\begin{document}

\title{A General Theory of Oscillon Dynamics}

\author{Marcelo Gleiser}
\email{gleiser@dartmouth.edu}

\author{David Sicilia}
\email{davidovich@dartmouth.edu}

\affiliation{Department of Physics and Astronomy, Dartmouth College,
Hanover, NH 03755, USA}

\date{\today}

\begin{abstract}
We present a comprehensive, nonperturbative analytical method to investigate the
dynamics  of time-dependent oscillating scalar field configurations. The method
is applied to oscillons in a $\phi^4$ Klein-Gordon model in two and three
spatial dimensions, yielding high accuracy results in the characterization of
all aspects of the complex oscillon dynamics. In particular, we show how
oscillons can be interpreted as long-lived perturbations about an attractor in
field configuration space. By investigating their radiation rate as they
approach the attractor, we obtain an accurate estimate of their lifetimes in
$d=3$ and explain why they seem to be perturbatively stable in $d=2$, where $d$
is the number of spatial dimensions.
\end{abstract}

\maketitle

\section{Introduction}

Nonlinear field theories contain a large number of localized solutions that
display a rich array of properties \cite{Rajamaran}. Of particular interest are
those that are static and stable, that is, that retain their spatial
profile as they move across space or scatter with each other, as is the case of
sine-Gordon solitons. The details of the solitonic configurations are, of
course, sensitive to the dimensionality of space and to the nature of the field
interactions. Long ago, Derrick has shown that, in the case of models with just
a real scalar field, no static solitonic configurations can exist in more than
one spatial dimension \cite{Derrick}. Given that most models of interest in high
energy  physics involve more complicated fields in three spatial dimensions,
this restriction was somewhat frustrating. Fortunately, the subsequent
exploration of a variety of
models led to a plethora of static, nonperturbative, localized field
configurations. Examples include topological defects, solutions of
models usually involving gauge fields that owe their stability to the nontrivial
topology of the vacuum, such as strings and monopoles \cite{Vilenkin}, and the
so-called nontopological solitons, solutions of models where a conserved global
charge is trapped inside a finite region of space due to a mass gap condition,
such as $Q$-balls \cite{Coleman} and the models with a real and a complex scalar
field of Friedberg, Lee, and Sirlin \cite{NTS}.

In the mid-nineties \cite{osc_glei1}, a new class of localized nonperturbative
solution began to be explored in detail, after being proposed earlier
\cite{Bogoliubov}. Named oscillons, such long-lived solutions have the
distinctive and
counter-intuitive feature of being time-dependent. In spite of this, the
nonlinear interactions act to preserve the localization of the energy, which
remains approximately constant for a
surprisingly long time \cite{osc_glei2}. During the past few years, oscillons
have attracted much interest. Their properties were explored in two
\cite{osc_2d} and higher \cite{osc_ddim} spatial
dimensions, in the presence of gauge fields \cite{osc_U1}, in the
standard model of particle physics \cite{osc_EW}, and in a simple cosmological
setting \cite{osc_cosm}. There have also been detailed attempts at understanding
some properties of oscillon-related configurations (typically with
small-amplitude oscillations), including their longevity, using perturbative
techniques \cite{osc_hungary}. On the other hand, a treatment explaining the
remarkable longevity of oscillons in models related to spontaneous
symmetry breaking, and thus of obvious interest in particle physics and
cosmology, has been lacking. The situation was partially remedied recently, when
we published a preliminary treatment of the problem \cite{osc_prl}. In the
present work, we greatly extend the range of our dynamical theory of oscillons
in scalar field models, include the details of many key derivations and
demonstrate its accuracy in reproducing numerical results. Our approach is
general enough to be extended to different scalar field models that exhibit
long-lived,
time-dependent localized configurations.

\section{Linear vs. Nonlinear Dynamics and the Oscillon Mass Gap}

In order to introduce some of the basic quantities needed for our theory, it is
instructive to start by reviewing some of the main properties of relativistic
oscillons. We will do so in the context of a simple $\phi^4$ model with a
symmetric double-well potential, as this is also the main focus of the present
work. To begin, consider the Lagrangian for a spherically-symmetric, real scalar
field in $d$-spatial dimensions,
\begin{equation}
\label{Lager1}
L = c_{d}\int r^{d-1}dr\left[\frac{1}{2}\dot \phi^2 -
\frac{1}{2}\left(\frac{\partial\phi}{\partial r}\right)^2 - V(\phi)\right],
\end{equation}
where $V(\phi) = m^2\phi^2$, and $c_d = 2\pi^{d/2}/\Gamma(d/2)$ is the
unit-sphere
volume in $d$ dimensions. Quantities are scaled to be dimensionless as follows:
$\phi = m^{(d-1)/2}\phi_0$ and $r^{\mu}=r_0^{\mu}/m$. We will henceforth only
use dimensionless variables, dropping the subscript ``$0$''.

We start by investigating the linear theory so that we can more easily contrast
it with nonlinear models that give rise to oscillons. Since oscillons have been
shown to maintain their approximate Gaussian-shaped spatial profiles during
their lifetimes, we will write the scalar field as 
\begin{equation}
\label{ansatz}
\phi(r, t) = A(t)P(r; R) = A(t)e^{-r^2/R^2}.
\end{equation}
Here, $A(t)$ is the time-dependent amplitude of the configuration and $P(r;R)$
its spatial profile, which is parameterized by the radial extension $R$.

\subsection{Linear Dynamics}

As shown in Ref. \cite{osc_glei2}, the $d=3$ linear theory with a
Gaussian-profile initial condition has the solution,
\begin{equation}
\label{eigenfunction1}
\phi(r, t) = \frac{R^3}{2}\frac{A}{\sqrt[]{\pi}}\int_{0}^{\infty} k e^{-R^2 k^2
/4}\frac{\sin (k r)}{r} \cos\left(\omega t\right)dk,
\end{equation}
where $A$ is an arbitrary initial amplitude and the dispersion relation is
$\omega = (k^2 + \omega_{\rm{mass}}^2)^{\frac{1}{2}}$, where the mass frequency
$\omega_{\rm{mass}} = \sqrt{2}$.  To calculate $\Gamma_{\rm lin}$, the decay
width associated with the above solution, we recall that in \cite{osc_glei2} it
was also shown that Eq. \ref{eigenfunction1} can be approximately integrated to
obtain (at $r = 0$, the configuration's maximum amplitude)
\begin{equation}
\label{philinear1}
\phi(0, t) = \frac{A_{0}}{\left(1+\frac{2
t^2}{R^4}\right)^{\frac{3}{4}}}\cos\left(\sqrt{2}t +
\frac{3}{2}\tan^{-1}\left[\frac{\sqrt{2}t}{R^2}\right]\right),
\end{equation}
whose envelope of oscillation is given by $\phi(0, t) = A_{0}/(1+2
t^2/R^4)^{3/4}$, which reaches $1/e$ of its initial value in a time given by
$\mathcal{T}_{\rm{linear}} \simeq .836 \omega_{\rm{mass}}R^2$.  This yields the
linear decay width $\Gamma_{\rm lin}$,
\begin{equation}
\label{lineargamma}
\frac{1}{2}\Gamma_{\rm lin} = \frac{1}{\mathcal{T}_{\rm{linear}}} \simeq
\frac{1.196}{\omega_{\rm{mass}}R^2} \simeq \frac{.846}{R^2}.
\end{equation}

In the linear theory, any initial configuration--or excitation above the
vacuum--will quickly decay by emitting radiation. The key difference between the
linear and the oscillon-supporting nonlinear models is that, in the latter case,
the decay modes are strongly suppressed. It is this suppression that gives rise
to the oscillon's remarkable longevity. Our goal in this paper is to make this
statement quantitatively precise. To obtain the linear radiation distribution,
which we denote by $b(\omega)$, we simply take the $k$-space representation $k
e^{-R^2 k^2
/4}$ of the Gaussian in Eq. \ref{eigenfunction1} and express it in terms of
$\omega$ using the dispersion relation $\omega = (k^2 + 2)^{\frac{1}{2}}$:
\begin{align}
\label{Bomega}
b(\omega) &= k[\omega] e^{-R^2 k[\omega]^2 /4} = \left(\omega^2 -
2\right)^{\frac{1}{2}}e^{-R^2(\omega^2-2)/4}.
\end{align}
$b(\omega)$ is a lopsided distribution with frequencies above
$\omega_{\rm{mass}} = \sqrt{2}$ and peaked at $\omega_{\rm max} =
(2+2/R^2)^{1/2}$.  Now define $\omega_{\rm{left}}$ and $\omega_{\rm right}$ to
be the frequencies where the distribution $b(\omega)$ rises to half of its peak
value.  A straightforward calculation gives
$\omega_{\rm{left}} \simeq (2 + .203/R^2)^\frac{1}{2}$ and $\omega_{\rm{right}}
\simeq (2 + 7.38/R^2)^{\frac{1}{2}}$.  The radiation distribution is then
approximately centered on the frequency $\omega_{\rm{lin}}$, given by (see Fig.
\ref{fig:linearpeaks}),
\begin{align}
\label{omegalinear}
\omega_{\rm{lin}} &\equiv \frac{1}{2}\left(\omega_{\rm{left}} +
\omega_{\rm{right}}\right) \\ \nonumber
&\simeq \frac{1}{2}\left(2 + \frac{.203}{R^2}\right)^{\frac{1}{2}} +
\frac{1}{2}\left(2 + \frac{7.38}{R^2}\right)^{\frac{1}{2}},
\end{align}
which we take to represent the dominant linear radiation frequency.

\subsection{Nonlinear Dynamics and Decay Rate}

\begin{figure}
\includegraphics[width=.45\textwidth,height=2.3in]{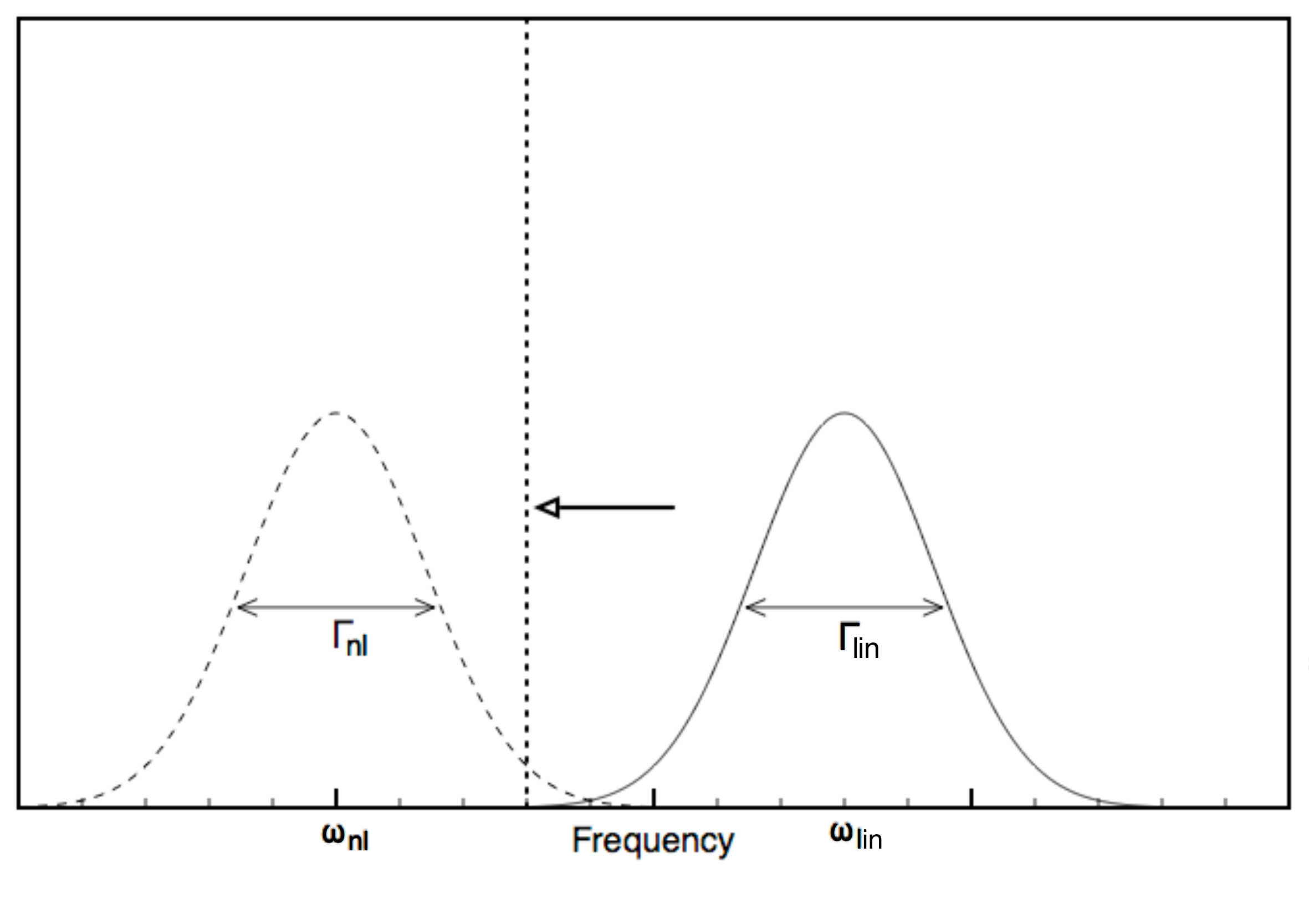}
\renewcommand{\baselinestretch}{1.0}
\caption[linearpeaks]
{Schematic description of the linear radiation peak centered on
$\omega_{\rm{lin}}$ being shifted to the left by the presence of nonlinearities. 
An oscillon will form if the peak is shifted far enough into the nonlinear
region such that it no longer overlaps significantly with the
linear peak.}
\label{fig:linearpeaks}
\end{figure}

Imagine now that one or more nonlinear terms are added to the linear potential
and that the field is again initialized with the same localized Gaussian
perturbation. 
Roughly speaking, the nonlinearities will shift the dominant linear oscillation
frequency $\omega_{\rm{lin}}$, to a new value, denoted by $\omega_{\rm{nl}}$. 
If the
added terms serve to decrease the curvature of the potential, then
$\omega_{\rm{nl}} <
\omega_{\rm{lin}}$.  Such a situation is depicted qualitatively in Fig.
\ref{fig:linearpeaks}, where the arrow indicates how the shift in frequencies
occurs.

When nonlinearities are efficient enough that $\omega_{\rm{nl}}$
is lowered substantially below $\omega_{\rm{lin}}$, an oscillon may form.  The
reason for this is the following.  As the initial field configuration begins to
oscillate, it will attempt to emit small-amplitude radiation waves in an effort
to dissipate its energy.  However, if $\omega_{\rm{nl}}$ is sufficiently less
than $\omega_{\rm{lin}}$, the bulk of the frequency components composing the
oscillation will be unable to excite small-amplitude radiation waves (since the
configuration can only radiate appreciably in the frequency range
$\omega_{\rm{left}} < \omega < \omega_{\rm{right}}$).  This condition leads to
the stabilization mechanism responsible for the formation of oscillons.

It then follows that the oscillon will enter the nonlinear regime if the
two peaks in Fig. \ref{fig:linearpeaks} do not significantly overlap. 
Mathematically, the
right ``edge'' of the nonlinear peak, given by $\omega_{\rm{nl}} +
\frac{1}{2}\Gamma_{\rm nl}$, must be less than the left ``edge'' of the linear
peak, given by
$\omega_{\rm{lin}} - \frac{1}{2}\Gamma_{\rm{lin}}$.  Therefore, we have
$\omega_{\rm{nl}} +
\frac{1}{2}\Gamma_{\rm nl} < \omega_{\rm{lin}} - \frac{1}{2}\Gamma_{\rm lin}$
which, by defining
$\omega_{\rm{gap}} \equiv \omega_{\rm{lin}} - \omega_{\rm{nl}}$ as the frequency
gap between the linear and nonlinear peaks, becomes
\begin{equation}
\label{condition1}
\omega_{\rm{gap}} > \frac{1}{2}\left (\Gamma_{\rm nl} + \Gamma_{\rm lin}\right).
\end{equation}
Since nonlinearities tend to increase a configuration's lifetime and thus
decrease its
decay rate, it follows that $0 \leq \Gamma_{\rm nl} \leq \Gamma_{\rm lin}$. 
One can thus state that if  
\begin{equation}
\label{condition}
\omega_{\rm{gap}} >  \Gamma_{\rm lin},
\end{equation}
then the configuration will be forced into the nonlinear regime.  
In other words, Eq. \ref{condition} is a \textit{necessary} condition
for the formation of an oscillon.  As we will soon see, if an oscillon is formed
and, during the course of its time
evolution, reaches a point where Eq. \ref{condition} is no longer satisfied, it
will cease to exist.  This implies that oscillons decay when
\begin{equation}
\label{oscildeath}
\omega_{\rm{gap}} = \Gamma_{\rm lin}.
\end{equation}

To obtain the nonlinear radiation frequency $\omega_{\rm{nl}}$ and thus
$\omega_{\rm gap}$, we substitute Eq. \ref{ansatz} into
Eq. \ref{Lager1} and integrate, giving
\begin{align}
\label{LagerA}
L &= \left(\frac{\pi}{2}\right)^{\frac{d}{2}}R^d\left[\frac{1}{2}\dot A^2 -
V(A_{\rm{max}})\right]; \nonumber \\
E &= \left(\frac{\pi}{2}\right)^{\frac{d}{2}}R^d V(A_{\rm{max}}),
\end{align}
where $V(A)$ now includes nonlinear terms and
$E$ is the energy which is found by taking the appropriate Legendre
transform of the Lagrangian and evaluating it at the upper turning point of an
oscillation, $A_{\rm{max}}$.  The oscillation frequency of the oscillon,
$\omega_{\rm{nl}}$, at a given time is given by
\begin{equation}
\label{freqint}
\frac{2\pi}{\omega_{\rm{nl}}} = \mathcal{T}_{\rm{osc}} =
\int_{0}^{\mathcal{T}_{\rm{osc}}}dt =
2\int_{A_{\rm{max}}}^{A_{\rm{min}}}\frac{dA}{\dot A},
\end{equation}
where $\dot A = [2E/c_{R} - 2V(A)]^{1/2}$, $c_{R} \equiv (\pi/2)^{d/2}R^d$, and
$A_{\rm{min}}$ is given by $V(A_{\rm{min}}) = V(A_{\rm{max}})$.

\subsection{The Attractor Point}

\begin{figure}
\includegraphics[width=.45\textwidth,height=2.3in]{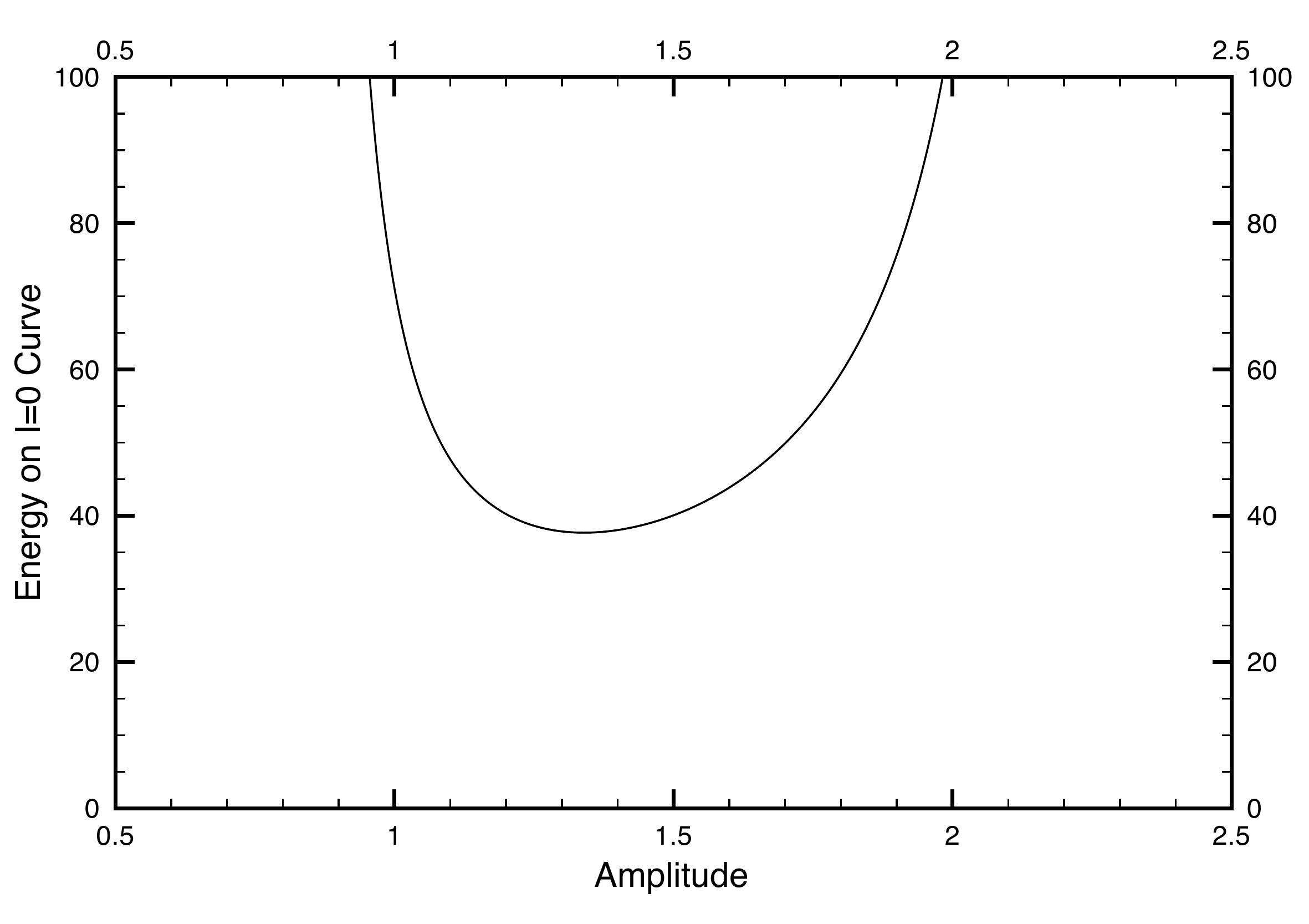}
\renewcommand{\baselinestretch}{1.0}
\caption[Einf]
{Minimum oscillon energy as a function of core amplitude for a double-well
potential in $d=3$. The minimum of this curve is the attractor point with
$E_{\infty} \simeq 37.69$.}
\label{fig:enmin}
\end{figure}

Results from numerical simulations suggest that there exists an attractor point
in
configuration space to which the oscillon tends.  It was noted in \cite{osc_prl}
that one can obtain the energy of this attractor point (in $\phi^4$ models) by
finding the minimum energy which has the property that the effective potential
$V(A)$ possesses at least one point for which $V''(A) \le 0$. In order to
compute the attractor point, it is easier to work within a specific model.
Choosing $V(A) = \phi^2-\phi^3 +\phi^4/4$, we obtain, using Eq. \ref{ansatz} and
integrating over all space,
\begin{equation}
\label{VA}
V(A) = \left(1 + \frac{d}{2 R^2}\right)A^2 -
\left(\frac{2}{3}\right)^{\frac{d}{2}}A^3 + \frac{A^4}{2^{\frac{d+4}{2}}},
\end{equation}
and,
\begin{equation}
\label{Istability}
V''(A) = \left(2+\frac{d}{R^2}\right) - 6\left(\frac{2}{3}\right)^{\frac{d}{2}}A
+ 3\frac{A^2}{2^{\frac{d}{2}}}.
\end{equation}
Equate this to zero and solve for $R$ as a function of $A$.  Then substitute the
result into Eq. \ref{LagerA} to eliminate $R$, yielding energy as a function
only of $A$.  This curve possesses a {\it minimum}, shown in Fig.
\ref{fig:enmin} for
$d=3$, the energy of which yields the correct attractor energy $E_{\infty}$ of
the oscillon, which has numerical values $E_{\infty} \simeq 4.44$ in $d=2$ and
$E_{\infty} \simeq 37.69$ in $d=3$.

\begin{figure}
\includegraphics[width=.45\textwidth,height=2.3in]{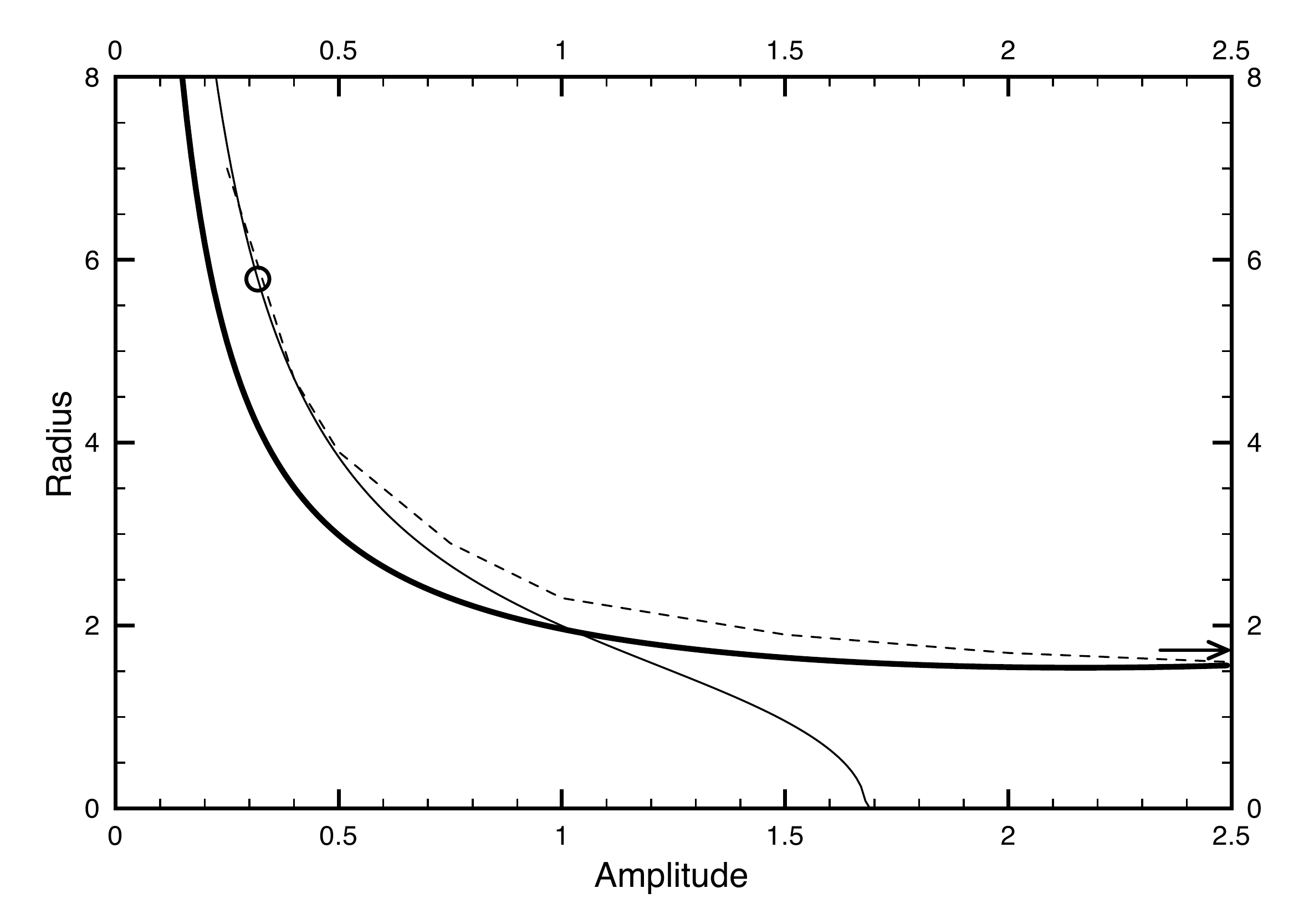}
\renewcommand{\baselinestretch}{1.0}
\caption[RA_deathD2]
{The thick solid line represents the locus of points satisfying the condition
$\omega_{\rm{gap}} = \Gamma_{\rm lin}$ in $d=2$ (``line of existence'')
calculated analytically.  The thin solid line represents the locus of points
that have the attractor energy $E \simeq 4.44$.  The dashed line is the
numerical minimum radius based on Gaussian initial configurations.  Oscillons
can only exist if their energy $E \ge E_{\infty}$ and if they have a core
amplitude and average radius lying above the line of existence.  This is why the
numerically measured minimum radius follows the line of existence for $A \gtrsim
1$ but follows the line of minimum energy for $A \lesssim 1$. The arrow
indicates the value of the (constant) minimum radius calculated in
\cite{osc_glei2}. The circle represents the location of the attractor point;
since it lies above the line of existence, oscillons will be absolutely stable
in this system.}
\label{fig:RA_deathD2}
\end{figure}

\begin{figure}
\includegraphics[width=.45\textwidth,height=2.3in]{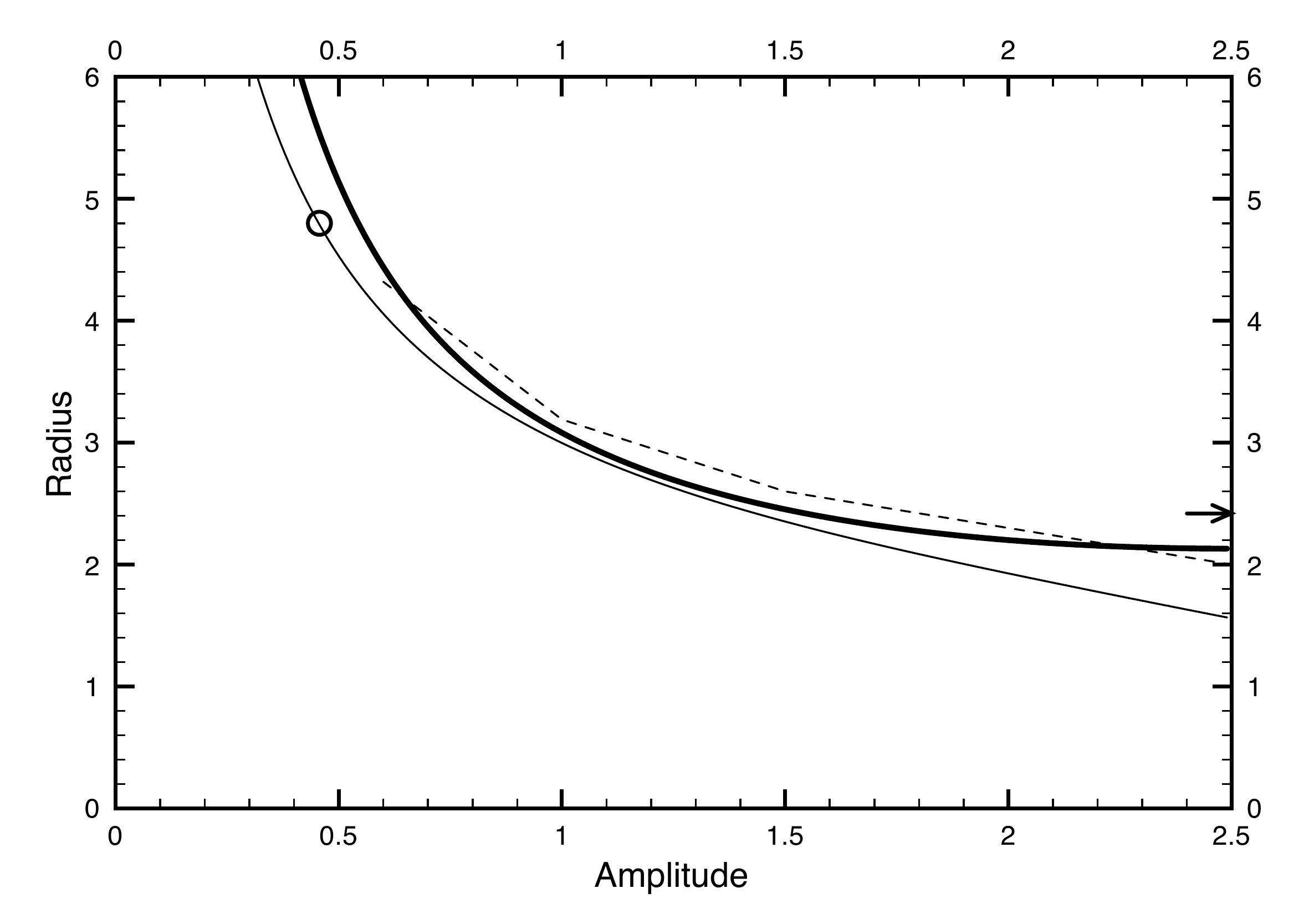}
\renewcommand{\baselinestretch}{1.0}
\caption[RA_death]
{The thick solid line represents the locus of points satisfying the condition
$\omega_{\rm{gap}} = \Gamma_{\rm lin}$ in $d=3$ (``line of existence'')
calculated analytically.  The thin solid line represents the locus of points
that have the attractor energy $E \simeq 37.69$.  The dashed line is the
numerical minimum radius based on Gaussian initial configurations. The arrow
indicates the value of the (constant) minimum radius calculated in
\cite{osc_glei2}. Oscillons can only exist if their core amplitude and average
radius are above the curve (wherein $E \ge E_{\infty}$ is automatically
satisfied).  The circle represents the location of the attractor point; since it
lies below the line of existence, oscillons will not be absolutely stable in
this system.}
\label{fig:RA_death}
\end{figure}

Given the energies calculated above, there is a locus of points in $(A, R)$
parameter space (the thinner solid lines in Figs. \ref{fig:RA_deathD2} and
\ref{fig:RA_death}) which possess these asymptotic values for the energy; one of
them is the attractor.  To locate the attractor point, we need to determine its
amplitude coordinate, denoted $A_{\infty}$.  In $d=2$, this is most easily
determined numerically to be $A_{\infty} \simeq .3$.  In $d=3$, we cannot
determine $A_{\infty}$ numerically since the oscillon decays before reaching it.
Therefore, in $d=3$, we must estimate $A_{\infty}$
analytically.  To do this, we choose the point (satisfying $E = E_{\infty}$)
which has $\omega_{\rm nl} \simeq
\omega_{\rm{mass}}$, even though,  in reality, its frequency is slightly
less than $\omega_{\rm{mass}}$ (never above it). As we will see, this
approximation
will suffice for our purposes.  From Eq. \ref{freqint}, the amplitude which
gives $\omega_{\rm{nl}}
\simeq \omega_{\rm{mass}}$ (in $d=3$) has numerical value $A_{\infty} \simeq
.456$.

Given the pair $(A_{\infty}, E_{\infty}) \simeq (.456, 37.69)$ in $d=3$ and
$(A_{\infty}, E_{\infty}) \simeq (.3, 4.44)$ in $d=2$, we can use Eq.
\ref{LagerA} to obtain $R_{\infty}$. We obtain $R_{\infty} \simeq 4.79$ for
$d=3$, and $R_{\infty} \simeq 5.77$ in $d=2$. The circles in Figs.
\ref{fig:RA_deathD2} ($d=3$) and \ref{fig:RA_death} ($d=2$) mark the locations
of the attractor
points.

Given Eqs. \ref{lineargamma}, \ref{omegalinear}, and \ref{freqint} (and their
equivalents in $d=2$ shown in Appendix A) we can also
calculate the quantities in Eq. \ref{condition}, that is, the oscillon existence
condition, as a function of the parameter
pair $(A, R)$.  In Figs. \ref{fig:RA_deathD2} ($d=2$) and \ref{fig:RA_death}
($d=3$), the thicker continuous lines represent
the locus of points satisfying $\omega_{\rm{gap}} = \Gamma_{\rm lin}$, defining
the boundary line between the region where oscillons may exist (above the line,
and provided that $E > E_{\infty}$)
and where they cannot exist (below the line).  We will often refer to this
boundary as the
``line of existence.'' As a test of the existence condition (Eq.
\ref{condition}),
we also plot the numerical result for the ``minimum radius'' as a function of
amplitude (dashed line), found by pinpointing the minimum initial radius which
causes a configuration to live longer than the linear decay time.  The arrows in
Figs. \ref{fig:RA_deathD2} and \ref{fig:RA_death} indicate the values of the
minimum radii calculated in \cite{osc_glei2} which possess no amplitude
dependence and thus provide only limited information.

As can be seen, the attractor point in $d=3$ lies {\it below} the line of
existence
curve, explaining the finite lifetimes of oscillons in that system: the
configurations decay before reaching the attractor point.  On the
other hand, the attractor point in $d=2$ lies {\it above} the curve, explaining
the
seemingly infinite oscillon lifetimes observed in numerical simulations. We can
thus interpret the oscillons as time-dependent perturbations about the attractor
point. Those in $d=3$ are unstable, albeit some can be extremely long-lived.
Those in $d=2$ are at least perturbatively stable.

In situations such as $d=3$, where oscillons eventually decay, it is interesting
to compute their lifetimes and how they depend on their radiation rate. In a
recent work, we presented the basic features of a method designed to do so
\cite{osc_prl}. In the next section, we develop the appropriate formalism in
detail, based on the overlap between the nonlinear and linear radiation spectra.
We point out that our formalism is, in principle, applicable to any
time-dependent scalar field configuration, offering a much-needed handle on how
to compute radiation rates of nonperturbative configurations in relativistic
scalar filed theories.

\section{Lifetime of Long-Lived Oscillons: General Theory}

In the situation that $\omega_{\rm{gap}} > \Gamma_{\rm lin}$ and an oscillon has
formed
(above the solid curve in Fig. \ref{fig:RA_death}) it will begin to radiate
small
amounts of energy.  In this section, we will derive a general equation governing
its
radiation rate so that, in the event that $\omega_{\rm{gap}} = \Gamma_{\rm lin}$
and the oscillon decays, we may calculate its lifetime.  As in the previous
section, our
general approach will be to compute the overlap between the nonlinear peak and
the linear peak.  Since we are assuming that $\omega_{\rm{gap}} > \Gamma_{\rm
lin} \gg \Gamma_{\rm{nl}}$ for a long-lived oscillon, this overlap will be
small; the amount by which it differs from zero will determine the
radiation rate.

\begin{figure}
\includegraphics[width=3in,height=2.5in]{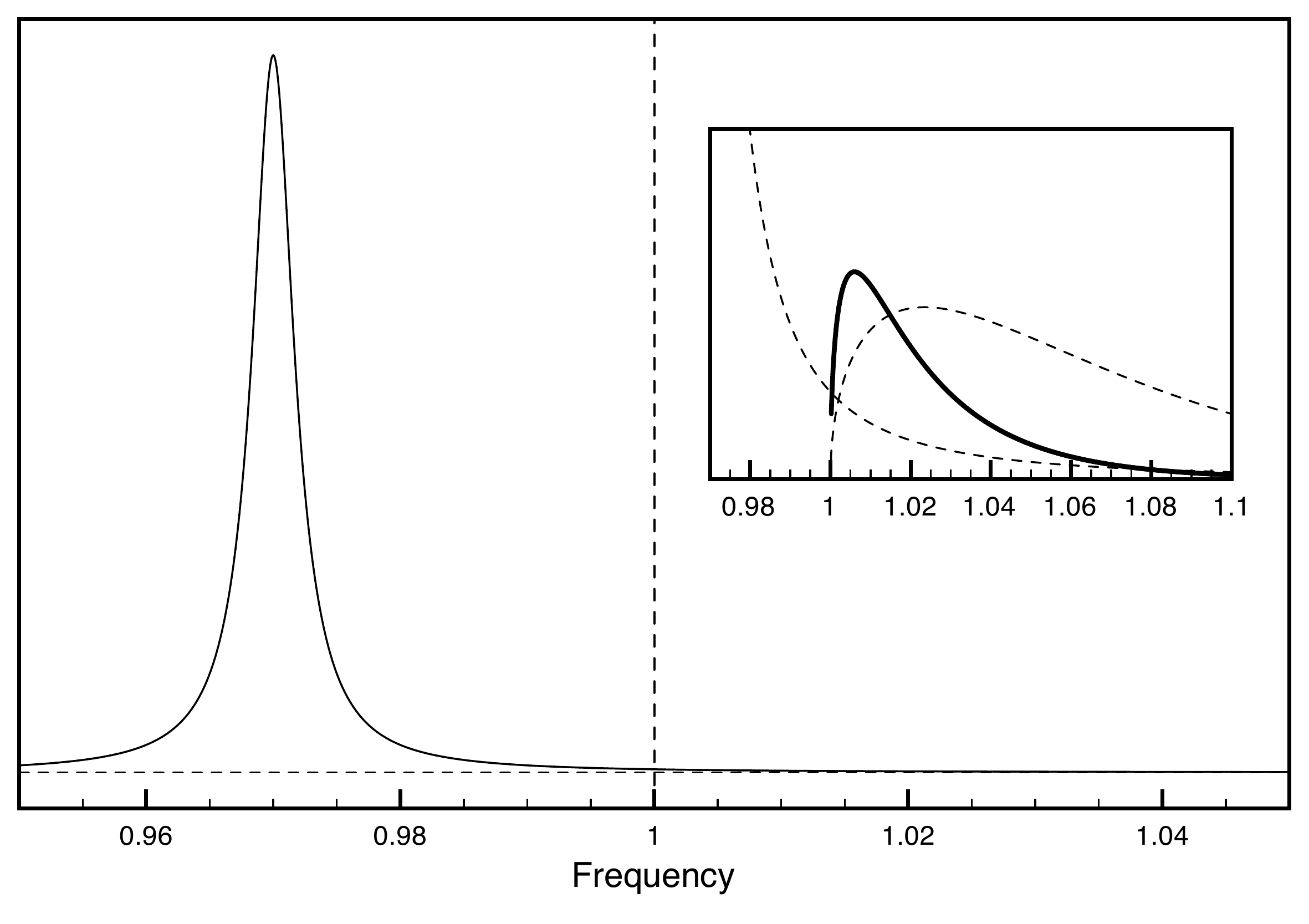}
\renewcommand{\baselinestretch}{1.0}
\caption[peaks]
{Schematic of an oscillon frequency distribution showing the tail penetrating
the radiation region. The graph is plotted in units of $\omega_{\rm mass}$. In
the inset we show a close-up view of  ${\cal F}(\omega)$ and $b(\omega)$,
respectively (dashed lines), and their product $\Omega(\omega)$ (solid line) for
the system given by Eq. \ref{Lager1} in $d=3$ for typical values of the various
parameters.  Note that the curves have been vertically scaled so that all are
visible on the same graph.}
\label{fig:peaks}
\end{figure}

In Fig. \ref{fig:peaks} we plot a possible frequency distribution for an
oscillon (the width of the peak centered on $\omega_{\rm{nl}}$ has been greatly
exaggerated for visibility) in units of the mass frequency ($\omega_{\rm mass} =
\sqrt{2}$). Note how the tail of the distribution ``leaks'' beyond the mass
frequency. It is this leakage that will determine the radiation-rate and thus
the decay rate of the oscillon. 

\subsection{The Long-Lived Oscillon Radiation Equation}

In order to compute the oscillon decay rate, we
model oscillons as spherically-symmetric objects whose radiation obeys a
distribution
(amplitude per unit frequency) $\Omega(\omega)$ which consists of a
narrow peak of width $\delta$ centered at some frequency $\omega_{\rm{rad}}$. 
The radiation flux $\Phi$ (energy per unit time per unit surface area) emitted
by such an object is
\begin{align}
\label{flux}
\Phi \equiv -\frac{\dot E}{S} &\simeq \rho v \simeq \frac{1}{2}\mathcal{A}^2
\omega_{\rm{rad}}^2\frac{\omega_{\rm{rad}}}{k_{\rm{rad}}} \\ \nonumber &\simeq
\frac{1}{2}\delta^2\Omega(\omega_{\rm{rad}})^2\frac{\omega_{\rm{rad}}^3}{k_{\rm{
rad}}}
\end{align}
where $S$ is the surface area of the oscillon, $\rho$ is the radiation-wave
energy density, $v = \omega_{\rm{rad}}/k_{\rm{rad}}$ is the phase velocity of
the wave, $k_{\rm{rad}}$ is the wave number, and $\mathcal{A}$ is the amplitude
of the radiation wave, which is given by
\begin{equation}
\label{radamp}
\mathcal{A} = \int_{\omega_{\rm{mass}}}^{\infty} \Omega(\omega)d\omega \simeq
\Omega(\omega_{\rm{rad}})\delta.
\end{equation}

The function $\Omega(\omega)$, which represents the amplitude per unit frequency
of the radiation wave, is simply determined by the ``overlap'' between
the oscillon and the linear radiation peaks.  Taking $\Omega(\omega)$ (the
\textit{overlap function}) to be the product of the nonlinear peak and the
linear radiation distribution $b(\omega)$ obtained in the previous section, we
have
\begin{align}
\label{overlap}
\Omega(\omega) &= \alpha \mathcal{F}(\omega) b(\omega); \\ \nonumber &\equiv
\mathcal{F}(\omega)\tilde b(\omega)
\end{align}
where $\mathcal{F}(\omega)$ is the nonlinear peak (Fourier transform of the
oscillon's core). $\alpha$ is a proportionality constant to be determined, and
$\tilde
b \equiv \alpha b(\omega)$.  The inset in Fig. \ref{fig:peaks} shows a typical
overlap function for an oscillon in $d=3$. (See Appendix C for details.)
In Appendix B we show that, in the tail,
\begin{equation}
\label{falpha}
\mathcal{F}(\omega) \simeq -\sqrt{\frac{2}{\pi}}\frac{\dot A}{1+\chi}(\omega -
\omega_{\rm{nl}})^{-2}.
\end{equation}

Combining Eqs. \ref{flux}, \ref{overlap}, and \ref{falpha} and letting $S =
c_{d}R^{d-1}$, we have
\begin{equation}
\label{AEeta}
\eta \frac{dE}{dt} + \left(\frac{dA}{dt}\right)^2 = 0,
\end{equation}
where $\eta$ is a time-dependent parameter given by
\begin{equation}
\label{eta}
\eta \equiv \frac{\pi(1+\chi)^2(\omega_{\rm{rad}} -
\omega_{\rm{nl}})^4k_{\rm{rad}}}{c_{d}R^{d-1}\omega_{\rm{rad}}^3\delta^2 \tilde
b(\omega_{\rm{rad}})^2}.
\end{equation}
Eq. \ref{AEeta} is a differential equation which must be satisfied by a
long-lived oscillon. In Appendix C we compute both $\alpha$ and $\delta$ in
general and in the context of the model of Eq. \ref{Lager1}.

\subsection{Integration of the Long-Lived Oscillon Radiation Equation}

In this section, we will attempt to integrate Eq. \ref{AEeta} to obtain the
oscillon energy as a function of time.  This process is not straightforward
since Eq. \ref{AEeta}
contains derivatives of two different quantities (amplitude and energy) and the
parameter $\eta$ possesses a complicated time dependence which is not known. 
However, in Appendix D we develop a simple method to solve this problem, based
on the assumption that the timescale associated with the oscillon's loss of
energy is closely related to the timescales associated with the rates of change
of all other oscillon parameters. The result of our approach is Eq. \ref{exp2},
which we will employ below.

First, write Eq. \ref{exp2} for the cases $X = A$ and $X = \eta$ and substitute
into Eq. \ref{AEeta}, obtaining,
\begin{equation}
\label{stableexp}
\dot E = \frac{-1}{\rho_{A}^2 \gamma_{A}^2}\left[\eta_{\infty} +
\gamma_{\eta}(E-E_{\infty})^{\rho_{\eta}}\right](E-E_{\infty})^{2-2\rho_A}.
\end{equation}
Now, consider the situation where $\eta_{\infty} \simeq 0$ (which is the case in
the system we are studying here since, in $d=3$, the attractor point satisfies
$\omega_{\rm{nl}} \simeq \omega_{\rm{mass}}$, leading to $\omega_{\rm{rad}}
\simeq \omega_{\rm{mass}}$, which causes $\eta$ to be zero there).  In this
situation, Eq. \ref{stableexp} is somewhat simplified:
\begin{equation}
\label{stablefinal}
\dot E = -\gamma_{\dot E}\left(E - E_{\infty}\right)^{\rho_{\dot E}},
\end{equation}
where the constant $\gamma_{\dot E}$ is given by
\begin{equation}
\label{gammaEdot}
\gamma_{\dot E} \equiv \frac{\gamma_{\eta}}{\rho_{A}^2 \gamma_{A}^2},
\end{equation}
and the exponent $\rho_{\dot E}$ is (using that $[\dot E]_{\infty} =
0$)
\begin{equation}
\label{rhorad}
\rho_{\dot E} = 2(1 - \rho_A) + \rho_{\eta}.
\end{equation}
Eq. \ref{stablefinal} is a constant-coefficient, ordinary differential
equation governing the oscillon energy as a function of time, and can be easily
integrated:
\begin{equation}
\label{Et}
E(t) = E_{\infty} + \frac{E_i - E_{\infty}}{\left[1 + \gamma_{\dot E}g(E_i -
E_{\infty})^{g} t \right]^\frac{1}{g}},
\end{equation}
where $g \equiv \rho_{\dot E} - 1$ and $E_{i}$ is the energy at $t = 0$.  Eq.
\ref{Et} is the energy of an oscillon as a function of time.

As shown in Fig. \ref{fig:RA_death}, in $d = 3$ an oscillon will always decay
before reaching
$E = E_{\infty}$.  The decay is quite sudden, a burst of scalar radiation. As
stated in section II, this occurs when
$\omega_{\rm{gap}} = \Gamma_{\rm lin}$; hence the decay energy, denoted
$E_{\rm{D}}$, is
given by
\begin{equation}
\label{deathenergy}
E_{\rm{D}} = E\vert_{[\omega_{\rm{gap}} = \Gamma_{\rm{lin}}]}.
\end{equation}

To calculate the lifetime, denoted $\mathcal{T}_{\rm{life}}$, which is defined
as the amount of time taken for the oscillon to decay from a sufficiently high
initial energy down to $E_{\rm{D}}$, we invert Eq. \ref{Et} to yield time as a
function of energy and evaluate at $E_{\rm{D}}$:
\begin{equation}
\label{life1}
t(E_{\rm{D}}) = \frac{1}{\gamma_{\dot
E}g}\left[\frac{1}{(E_{\rm{D}}-E_{\infty})^{g}} -
\frac{1}{(E_i-E_{\infty})^{g}}\right].
\end{equation}
When $E_i - E_{\infty} \gg E-E_{\infty}$, the function $t(E)$ tends to a
finite, maximum value leading to
\begin{equation}
\label{lifetime}
\mathcal{T}_{\rm{life}} = \frac{1}{\gamma_{\dot
E}g}\frac{1}{[E_{\rm{D}}-E_{\infty}]^{g}}.
\end{equation}
This means that, when the initial energy $E_{i}$ is much larger than the energy
$E_{\rm{D}}$ in question, the time it takes for $E(t)$ to fall from $E_{i}$ to
$E_{\rm{D}}$ becomes independent of the initial condition (i.e., from $E_i$).
One can then say that, in a restricted sense, the long-lived oscillon is
decoupled from initial conditions: if the necessary conditions for its existence
are satisfied, a variety of initial configurations will approach an oscillon.
Recent
studies that have observed the emergence of oscillons from stochastic initial
conditions after a fast quench offer strong support for this claim
\cite{osc_emerge}.

Eqs. \ref{deathenergy} and \ref{lifetime} together give the lifetime of an
oscillon and can be considered the main
results of this paper.  Before moving on, we note that, for a long-lived
oscillon,
$\Gamma_{\rm{nl}}$ is given by
\begin{align}
\label{gammanl}
\Gamma_{\rm{nl}} &\equiv \Gamma_{A} = -\frac{\dot A}{A - A_{\infty}} =
-\rho_A\frac{\dot E}{E-E_{\infty}} \\ \nonumber &= \rho_A\gamma_{\dot
E}(E-E_{\infty})^g,
\end{align}
where we've used Eqs. \ref{width}, \ref{exp03}, and \ref{stablefinal}.  This is
related to the lifetime by (combine Eqs. \ref{lifetime} and \ref{gammanl})
\begin{equation}
\label{lifewidth}
\mathcal{T}_{\rm{life}} = \left(\frac{\rho_A}{g}\right)
\Gamma_{\rm{nl}}^{-1}\biggl \vert_{E = E_{\rm{D}}}.
\end{equation}

\subsection{Sample Calculation: $\phi^{4}$ Klein-Gordon Field in $d=3$}

We now apply the above results to the system given by the Lagrangian in
Eq. \ref{Lager1} for $d=3$, supplemented by the nonlinear potential of Eq.
\ref{VA}.  We will begin with the existence condition shown in
Fig. \ref{fig:RA_death}.  In section II.c we calculated the coordinates of the
attractor point and obtained $(A_{\infty}, R_{\infty}) \simeq (.456, 4.79)$. 
Comparison with Fig. \ref{fig:RA_death} reveals that the attractor point
lies below the curve, and thus that $\omega_{\rm gap}/\Gamma_{\rm lin} < 1$
there.  As we noted before, stable oscillons will not exist in this
model.  However, there are still long-lived oscillons obtained by
initializing the field sufficiently far from
the attractor point.

\begin{figure}
\includegraphics[width=.45\textwidth,height=2.3in]{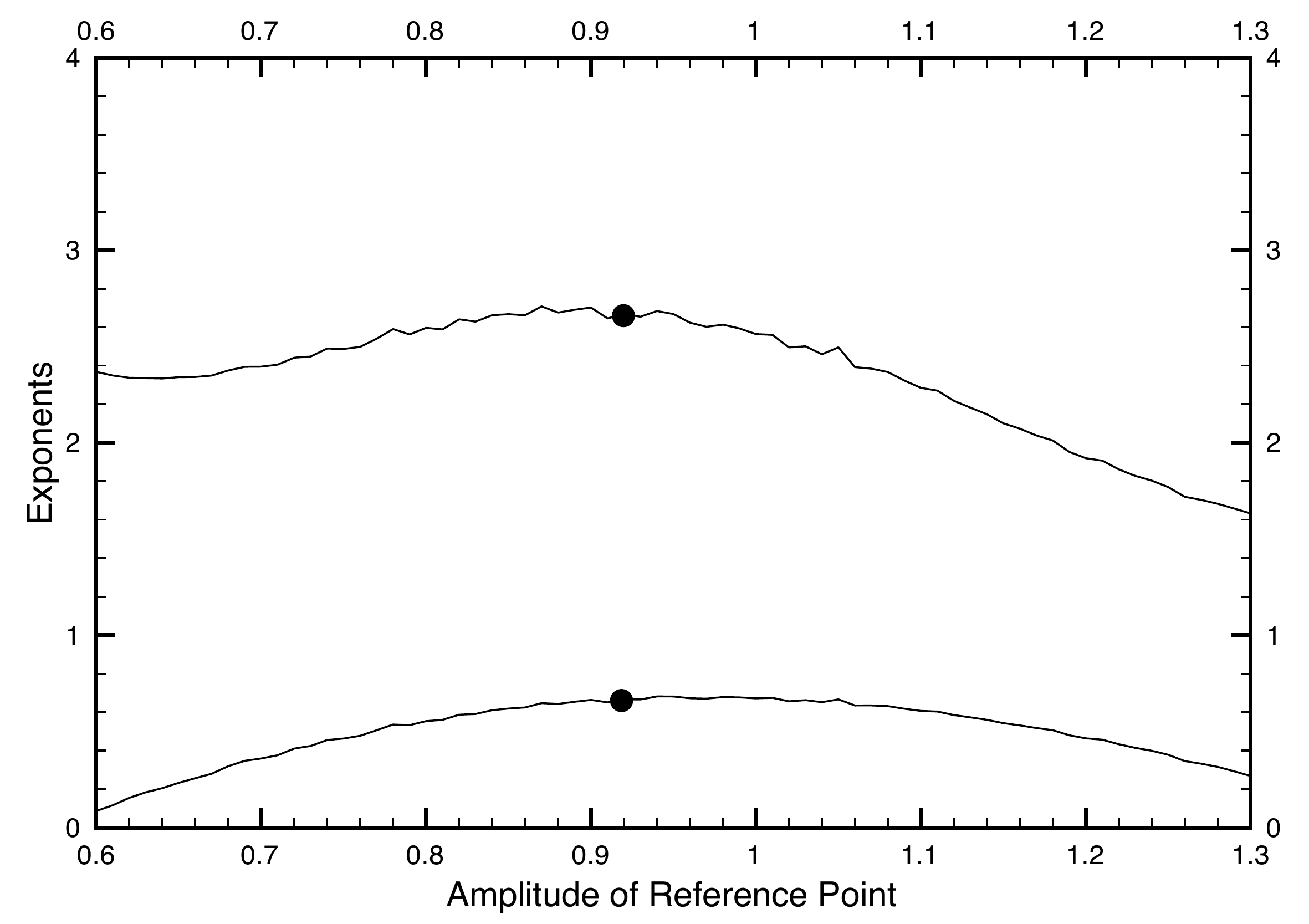}
\renewcommand{\baselinestretch}{1.0}
\caption[g_calc]
{Analytical calculation of $g = \rho_{\dot E} - 1$ (top curve)
and $\rho_A$ (bottom curve) using Eq. \ref{rhoXfinal3}, plotted against the
reference amplitude $A_r$ along the line of existence.  This
gives the spectrum of possible values of $g$ and $\rho_A$ across the various
oscillons in this system.  The dots mark the theoretical values of $g
\simeq 2.67$ and $\rho_A \simeq .66$ assumed by the longest-lived oscillon,
which has $A_r \simeq .92$ (thicker line in Fig. \ref{fig:trajectory}).}
\label{fig:g_calc}
\end{figure}

\begin{figure}
\includegraphics[width=.45\textwidth,height=2.3in]{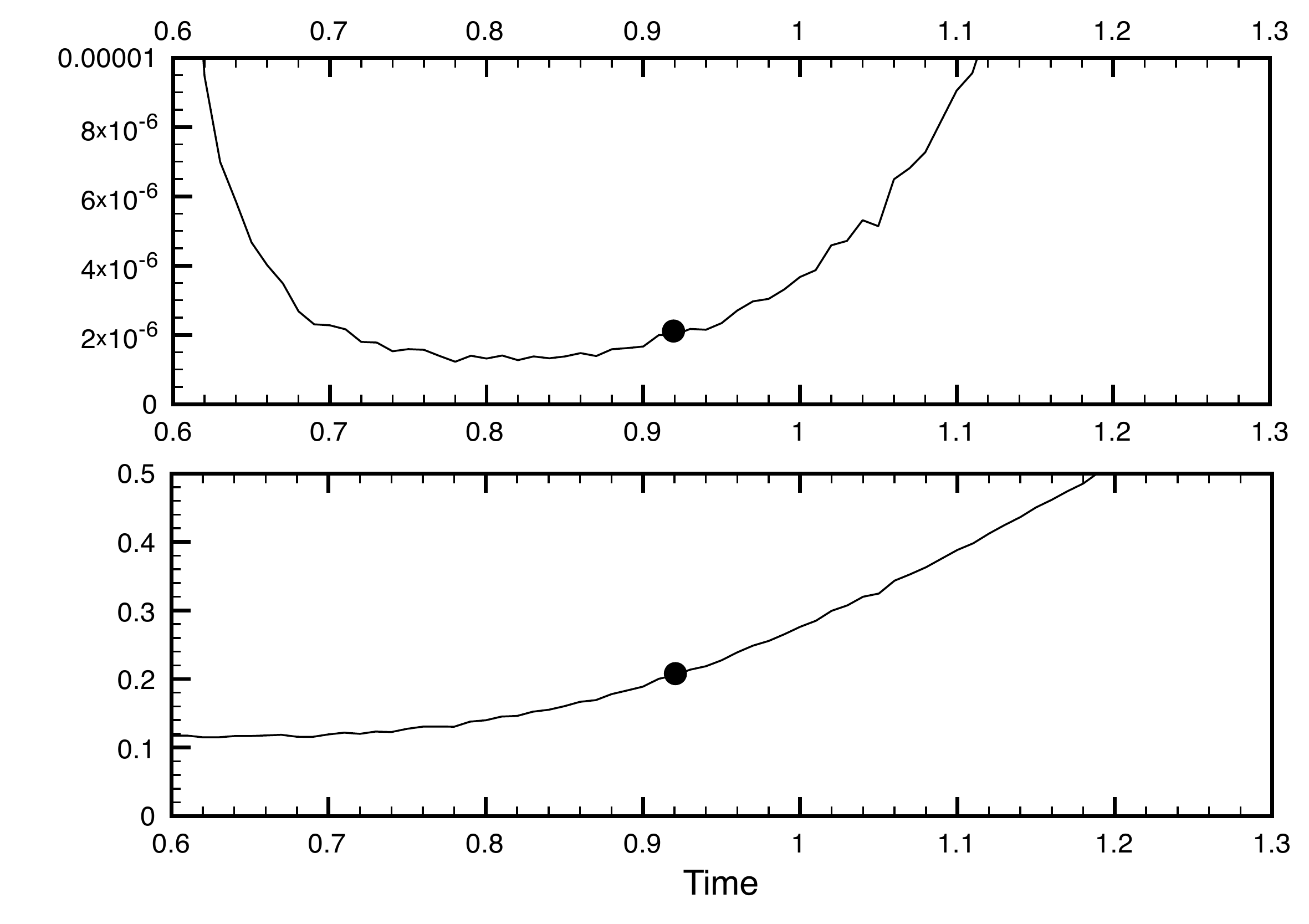}
\renewcommand{\baselinestretch}{1.0}
\caption[gammas]
{Analytical calculation of $\gamma_{\dot E}$ (top graph) and
$\gamma_A$ (bottom graph) using Eq. \ref{gammaX}, plotted against the reference
amplitude $A_r$ along the line of existence.  This gives the
spectrum of possible values of $\gamma_{\dot E}$ and $\gamma_A$ across the
various oscillons in this system.  The dots serve to mark the theoretical values
of $\gamma_{\dot E} \simeq 2.0 \times 10^{-6}$ and $\gamma_A \simeq .21$
assumed by the longest-lived oscillon, which has $A_r \simeq .92$ (thicker line
in Fig. \ref{fig:trajectory}).}
\label{fig:gammas}
\end{figure}

As these structures radiate energy, their amplitude $A$ and radius $R$ will
change in time. Hence, they will trace out trajectories in the $(A,R)$ plane of
Fig. \ref{fig:RA_death}, all of which will eventually intersect the line of
existence.  To calculate these trajectories, begin by writing Eq. \ref{exp2} for
$A(t)$ and $R(t)$, obtaining $[A(t) - A_{\infty}] =
\gamma_{A}[E(t)-E_{\infty}]^{\rho_{A}}$ and $[R(t) - R_{\infty}] =
\gamma_{R}[E(t)-E_{\infty}]^{\rho_{R}}$, respectively.  Substituting the first
into the second to eliminate the energy, we obtain
\begin{equation}
\label{trajectory2}
R(t) = R_{\infty} + \gamma_{RA}[A(t)-A_{\infty}]^{\rho_{RA}},
\end{equation}
where $\gamma_{RA} \equiv \gamma_R / \gamma_A^{\rho_{RA}}$ and $\rho_{RA} \equiv
\rho_R / \rho_A$.  Note that when $A = A_{\infty}$, $R = R_{\infty}$.

Since each possible trajectory will intersect the line of existence, every point
along the line of existence is a point along some trajectory. 
Thus, we can choose an arbitrary point on this line (which we will call the
``reference point'') and use its coordinates, labeled $(A_r,R_r)$, to calculate
all of the dynamical exponents ($\rho_A$, $\rho_R$, $\rho_{\dot E}$,
$\rho_{\omega}$, $\rho_\eta$, etc.) associated with the trajectory that
intersects that point.  Eq.
\ref{rhoXfinal2}, reproduced below, was evaluated numerically for $X=\rho_{\dot
E}$ and $X=\rho_A$, respectively:
\begin{equation}
\label{rhoXfinal3}
\rho_X = \frac{\left(2 + \frac{\partial \nu_\eta}{\partial
\nu_E}\right)\frac{\partial \nu_{X}}{\partial \nu_E} + \left(\frac{\partial
\nu_\eta}{\partial \nu_A} - 2\right)\frac{\partial \nu_{X}}{\partial
\nu_A}}{\left(2 + \frac{\partial \nu_\eta}{\partial \nu_E}\right)}.
\end{equation}
Fig. \ref{fig:g_calc} shows the results for the exponents
$g \equiv \rho_{\dot E} - 1$ (top curve) and $\rho_A$ (bottom curve) as a
function of $A_{r}$.

Given values for $\rho_X$, we can calculate $\gamma_X$ by evaluating each
side of Eq. \ref{exp2} at the reference point (wherein $X$ assumes the value
$X_r = X[A_r, R_r]$) and solving for $\gamma_X$, obtaining
\begin{equation}
\label{gammaX}
\gamma_X = \frac{X_r - X_{\infty}}{(E_r - E_{\infty})^{\rho_X}}.
\end{equation}
Fig. \ref{fig:gammas} shows the results of such a calculation for $\gamma_{\dot
E}$ (top graph) and $\gamma_A$ (bottom graph).

\begin{figure}
\includegraphics[width=.45\textwidth,height=2.3in]{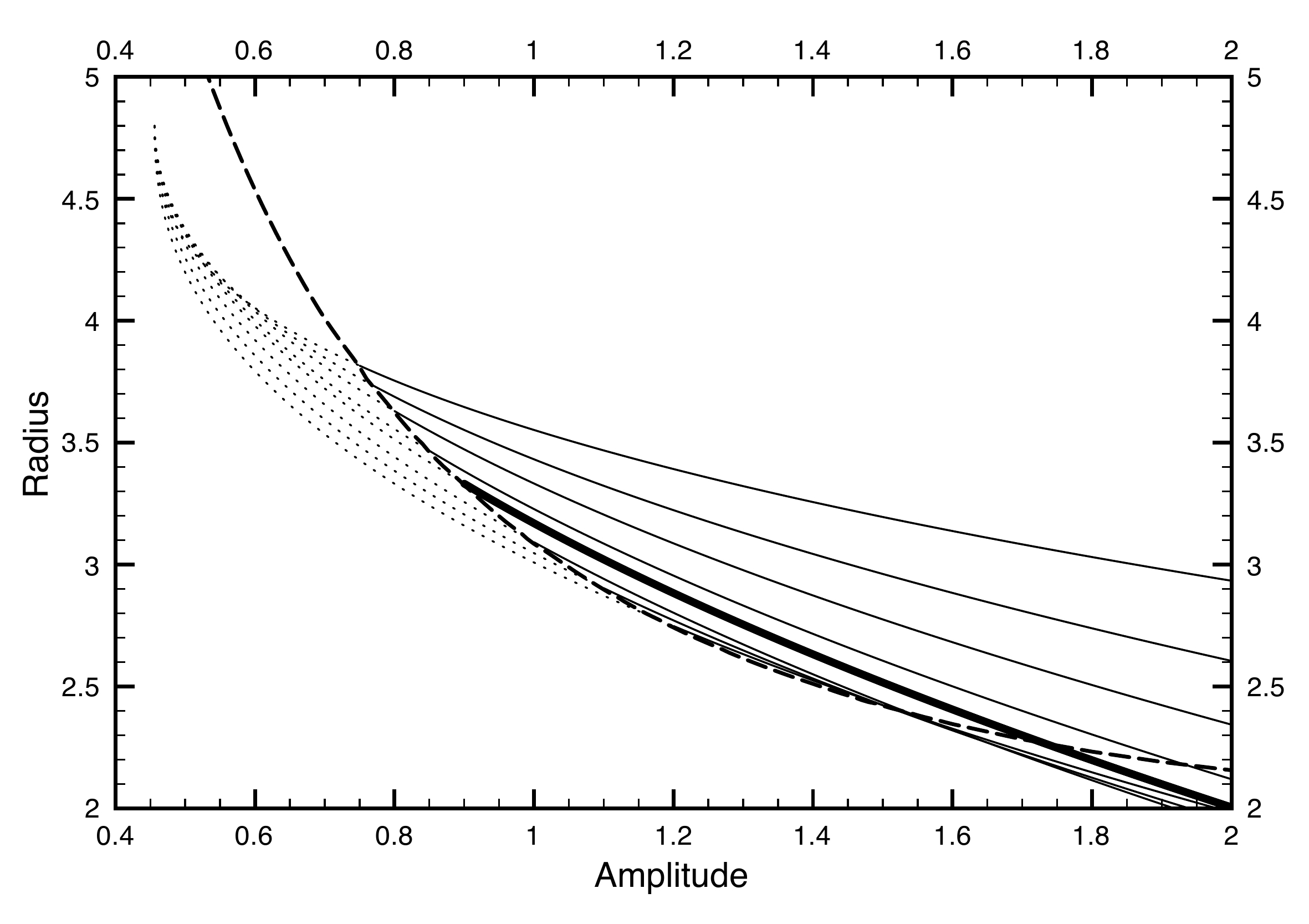}
\renewcommand{\baselinestretch}{1.0}
\caption[trajectory]
{Various oscillon trajectories.  Note that they all tend to the attractor
point but intersect the line of existence (dashed line) before doing so.  The
thicker trajectory marks the longest-lived oscillon
(it intersects the line of existence at $A_r \simeq .92$ and $R_r \simeq
3.25$).}
\label{fig:trajectory}
\end{figure}

Having computed $\rho_X$ and $\gamma_X$ for the parameters $A$ and $R$ allows us
to calculate the
coefficients in Eq. \ref{trajectory2} for each trajectory (i.e., compute Eqs.
\ref{rhoXfinal3} and \ref{gammaX} for various reference points along the line of
existence).  The result of this is shown in Fig. \ref{fig:trajectory}
for selected trajectories. Note that they all asymptotically tend toward the
attractor point but intersect the line of existence before doing so.

As an example, consider the trajectory which intersects the line of existence at
$A_r \simeq .92$ (this will be shown to correspond to the longest-lived oscillon
in the system of Eq. \ref{Lager1}).  From Figs. \ref{fig:g_calc} and
\ref{fig:gammas} we have $g \simeq 2.67$ and $\gamma_{\dot E} \simeq 2.0
\times 10^{-6}$.  If this oscillon were initiated at an energy of $E_i \simeq
82.5$ (which will be the case for the numerical simulation we will be comparing
to) then Eq. \ref{Et} for that oscillon becomes
\begin{align}
\label{Etlong}
E(t) &\simeq 37.69 + \frac{44.8}{\left[1+(.136)t\right]^{.375}},
\end{align}
which, for times $t \gtrsim 100$ is approximately
\begin{equation}
\label{Etlong2}
E(t) - E_{\infty} \simeq \frac{94.77}{t^{.375}}.
\end{equation}
Since the decay energy for this oscillon is $E_{\rm{D}} = E_r = E_r(A_r, R_r)
\simeq 41.0963$, Eq. \ref{lifetime} yields
\begin{align}
\label{lifelong}
\mathcal{T}_{\rm{life}} &\simeq \frac{1}{[2.0 \times
10^{-6}][2.67]}\frac{1}{[41.10-37.69]^{2.67}} \\ \nonumber
&\simeq 7100.
\end{align}

Given Eq. \ref{Etlong} we can write, for example, an expression for the oscillon
amplitude as a function of time.  From Figs. \ref{fig:g_calc} and
\ref{fig:gammas} we have $\rho_A \simeq .67$ and $\gamma_A \simeq .21$. 
Combining this information with $[A-A_{\infty}] = \gamma_A [E -
E_{\infty}]^{\rho_A}$ and Eq. \ref{Etlong} we have
\begin{equation}
\label{Along}
A(t) \simeq .456 + \frac{2.58}{\left[1+(.136)t\right]^{.249}}.
\end{equation}

\begin{figure}
\includegraphics[width=.45\textwidth,height=2.3in]{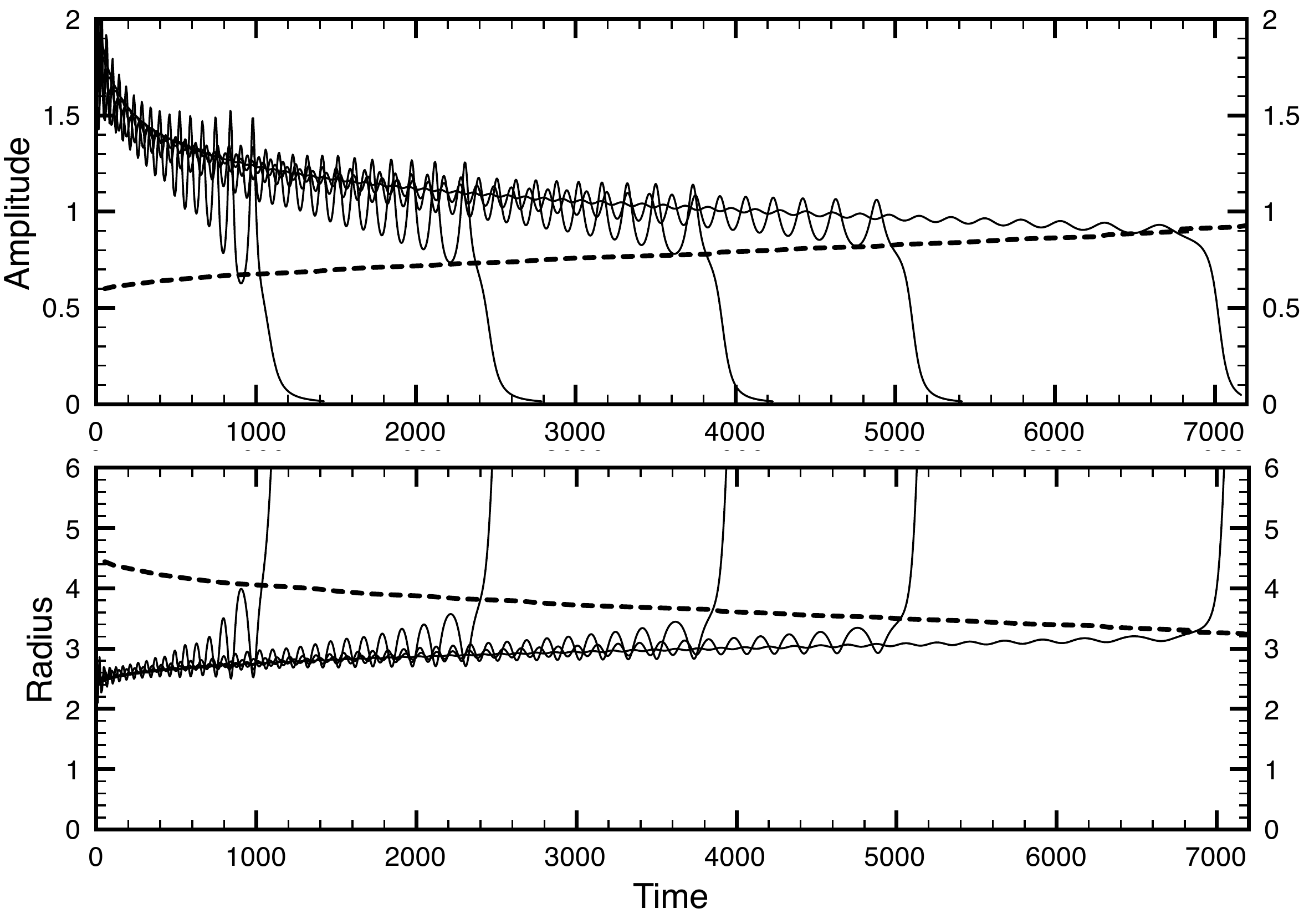}
\renewcommand{\baselinestretch}{1.0}
\caption[spectrum_amprad]
{Analytical results for critical values of the amplitude (top) and radius
(bottom) along the line of existence (dashed lines) as a function of lifetime
are plotted with several examples of short- and long-lived oscillons, all with
initial amplitudes $A_0=2$. From left to right, the initial radii for the
oscillons are $2.35$, $2.41$, $2.53$, $2.65$, and $2.86$. It is quite clear that
the oscillons decay as they cross the critical values computed analytically.}
\label{fig:spectrum_amprad}
\end{figure}

\begin{figure}
\includegraphics[width=.45\textwidth,height=2.3in]{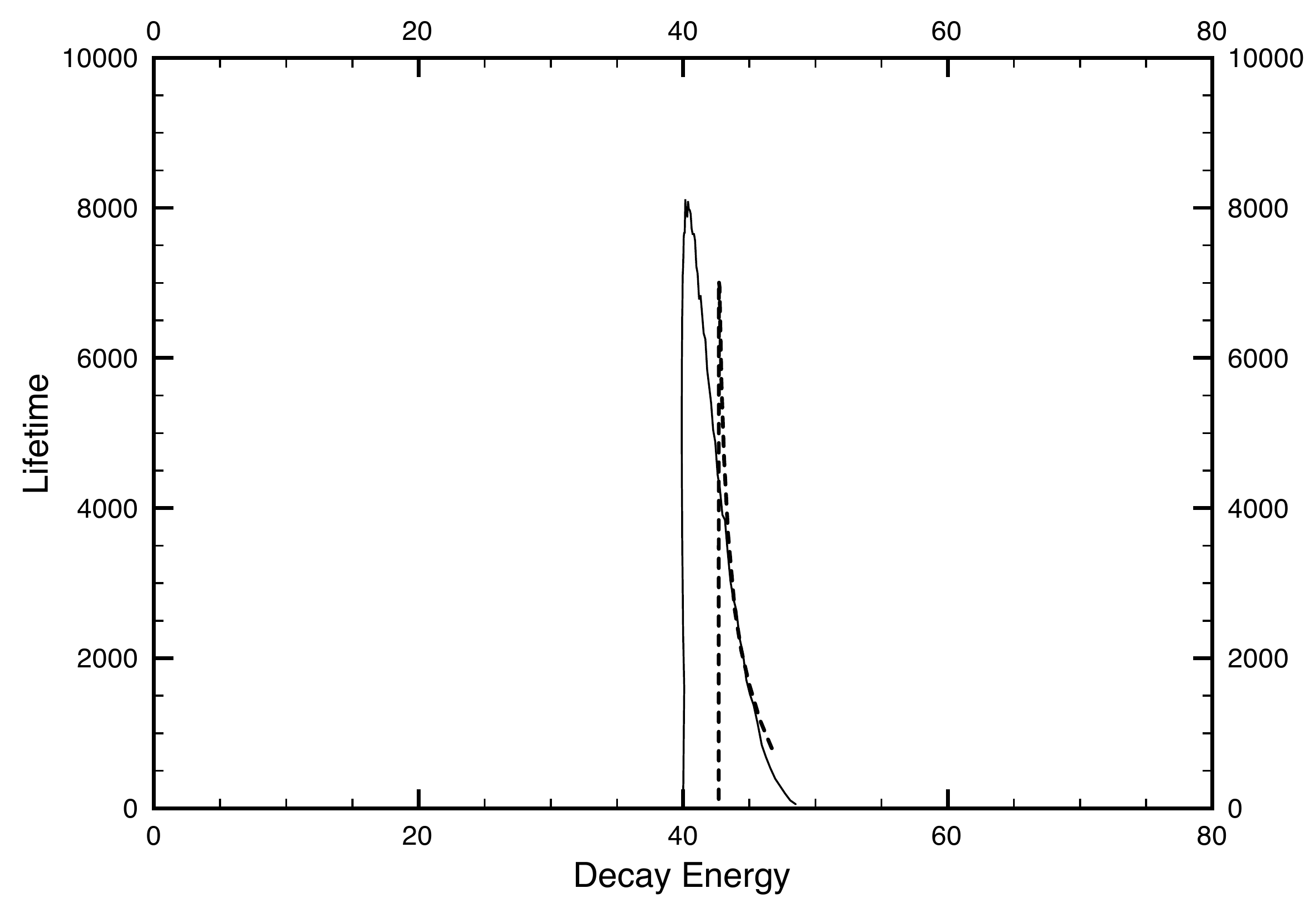}
\renewcommand{\baselinestretch}{1.0}
\caption[lifeenergy]
{Oscillon lifetimes vs. decay energy $E_{\rm{D}} = E_r$. 
Solid curve is theoretical, dashed line is numerical.  The theoretical curve
(with an error of $\sim 6\%$ in the horizontal positioning of the peak)
correctly predicts the shape of the distribution and that there exists a maximum
lifetime
in this system on the order of $\sim 10^4$.}
\label{fig:lifeenergy}
\end{figure}

Carrying out the above calculations for several trajectories along the
line of existence and plotting the lifetime vs. $A_r$,
$R_r$, and $E_r$, results in Figs. \ref{fig:spectrum_amprad} and
\ref{fig:lifeenergy}, respectively.  Fig. \ref{fig:spectrum_amprad} shows the
analytical decay amplitude and radius as a
function of lifetime (dashed lines) plotted against several long-lived and
short-lived oscillons, with excellent agreement: oscillons decay as they
cross the coordinates specified by the line of existence.  Fig.
\ref{fig:lifeenergy}
shows the computed lifetime as a function of decay energy ($E_D = E_r$),
reproducing the maximum lifetime on the order of $10^4$ which is characteristic
of
oscillons in this
system. The figure compares the analytical computation of lifetime (continuous
line) with the numerical results (dashed line). The small disparity in the
center of the peak (of order $\sim 6\%$) is probably due to the Gaussian {\it
ansatz} we use
to describe oscillon configurations.

\begin{figure}
\includegraphics[width=.45\textwidth,height=2.3in]{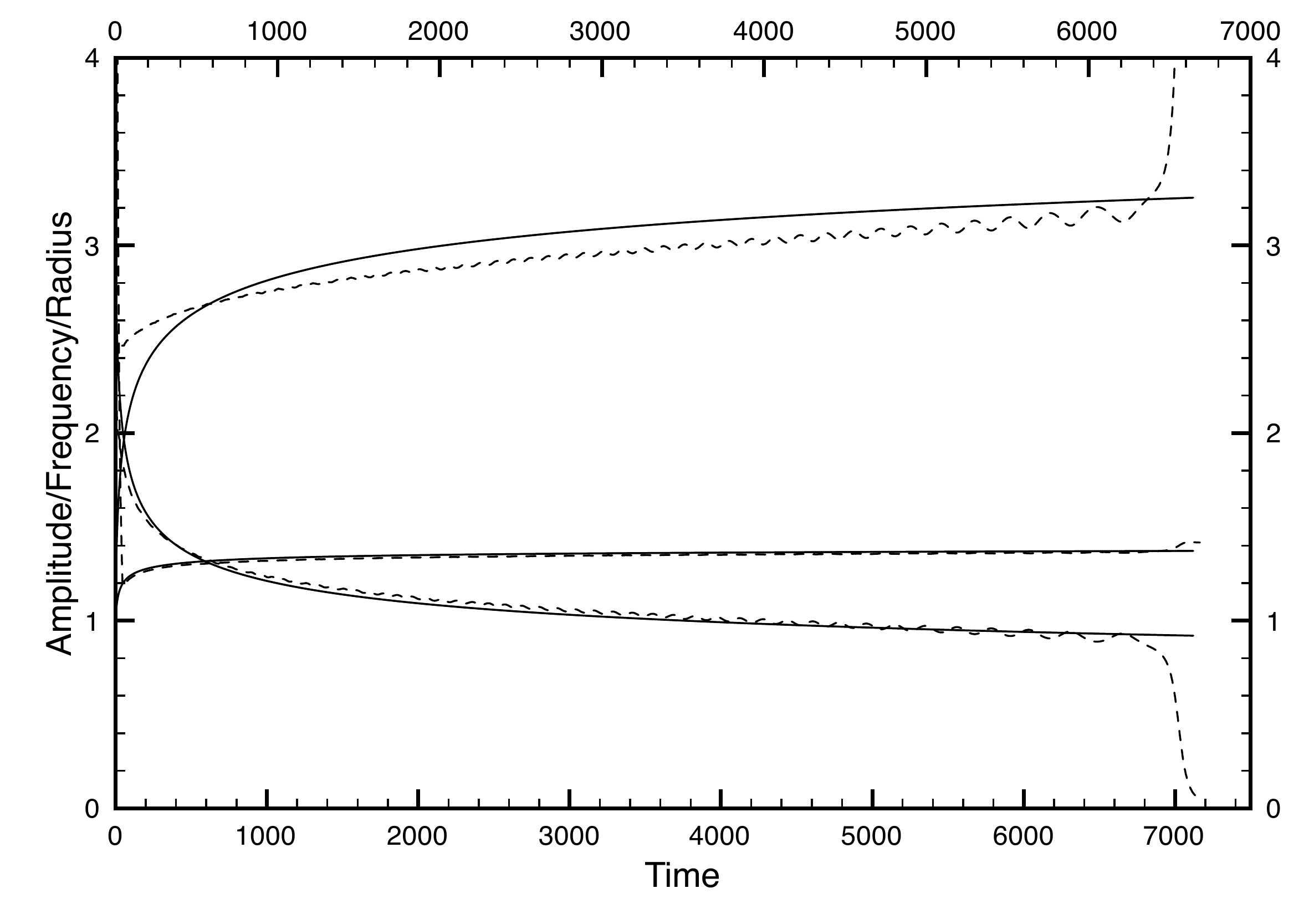}
\renewcommand{\baselinestretch}{1.0}
\caption[comp_afr]
{Comparison of theoretical (continuous line) vs. numerical (dashed line) radius
(top), frequency (middle) and amplitude (bottom) for an oscillon with initial
conditions ($A_0 = 2$, $R_0 =
2.86$), showing very good agreement [theoretical results are computed with
$(A_r,R_r) \simeq (.92, 3.25)]$.}
\label{fig:comp_afr}
\end{figure}

\begin{figure}
\includegraphics[width=.45\textwidth,height=2.3in]{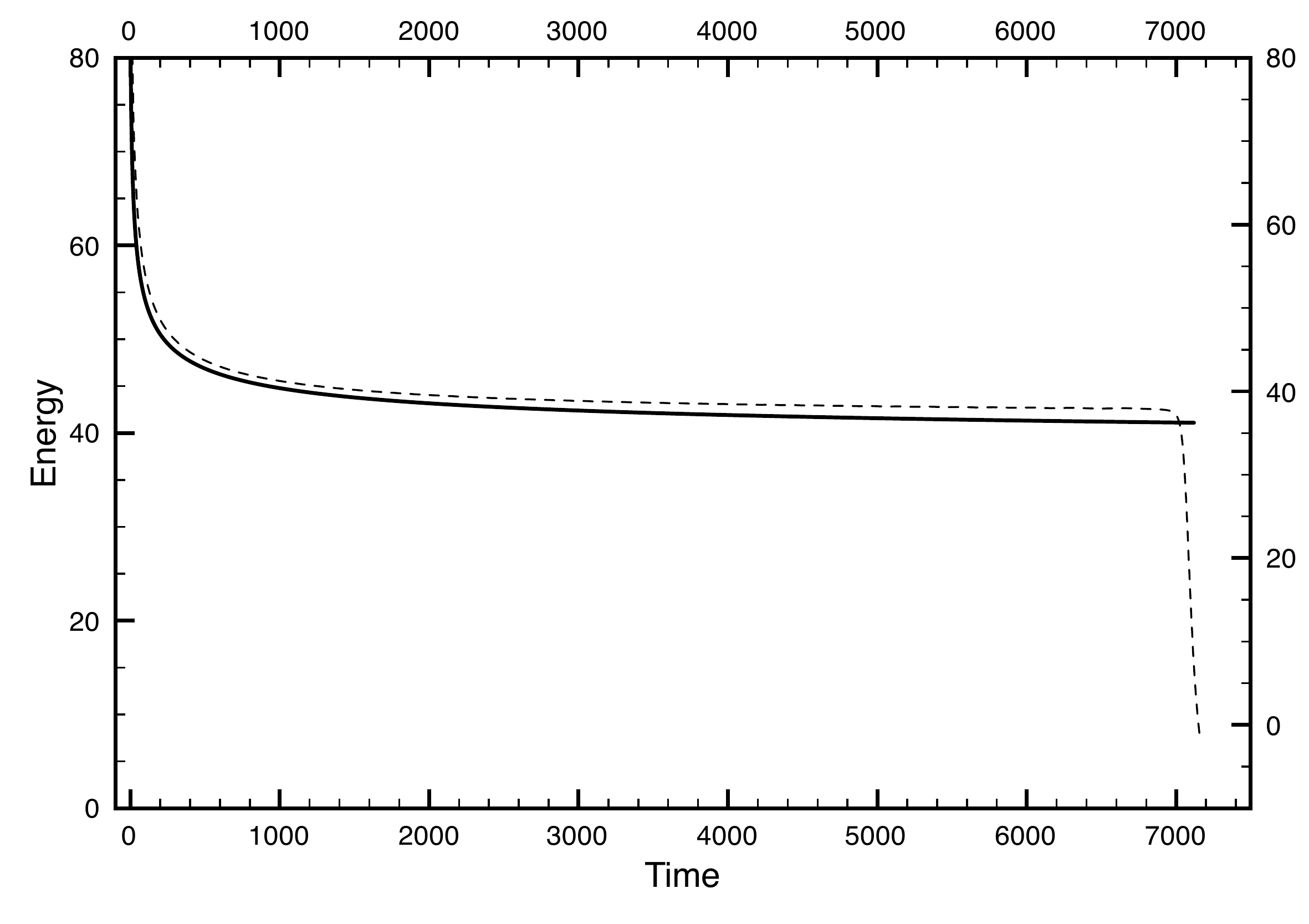}
\renewcommand{\baselinestretch}{1.0}
\caption[comp_nrg]
{Comparison of theoretical (continuous line) vs. numerical (dashed line) values
for the energy of the longest-lived oscillon, obtained with initial
conditions ($A_0 = 2$, $R_0 = 2.86$) showing excellent agreement [$(A_r,R_r)
\simeq (.92, 3.25)$].}
\label{fig:comp_nrg}
\end{figure}

\begin{figure}
\includegraphics[width=.45\textwidth,height=2.3in]{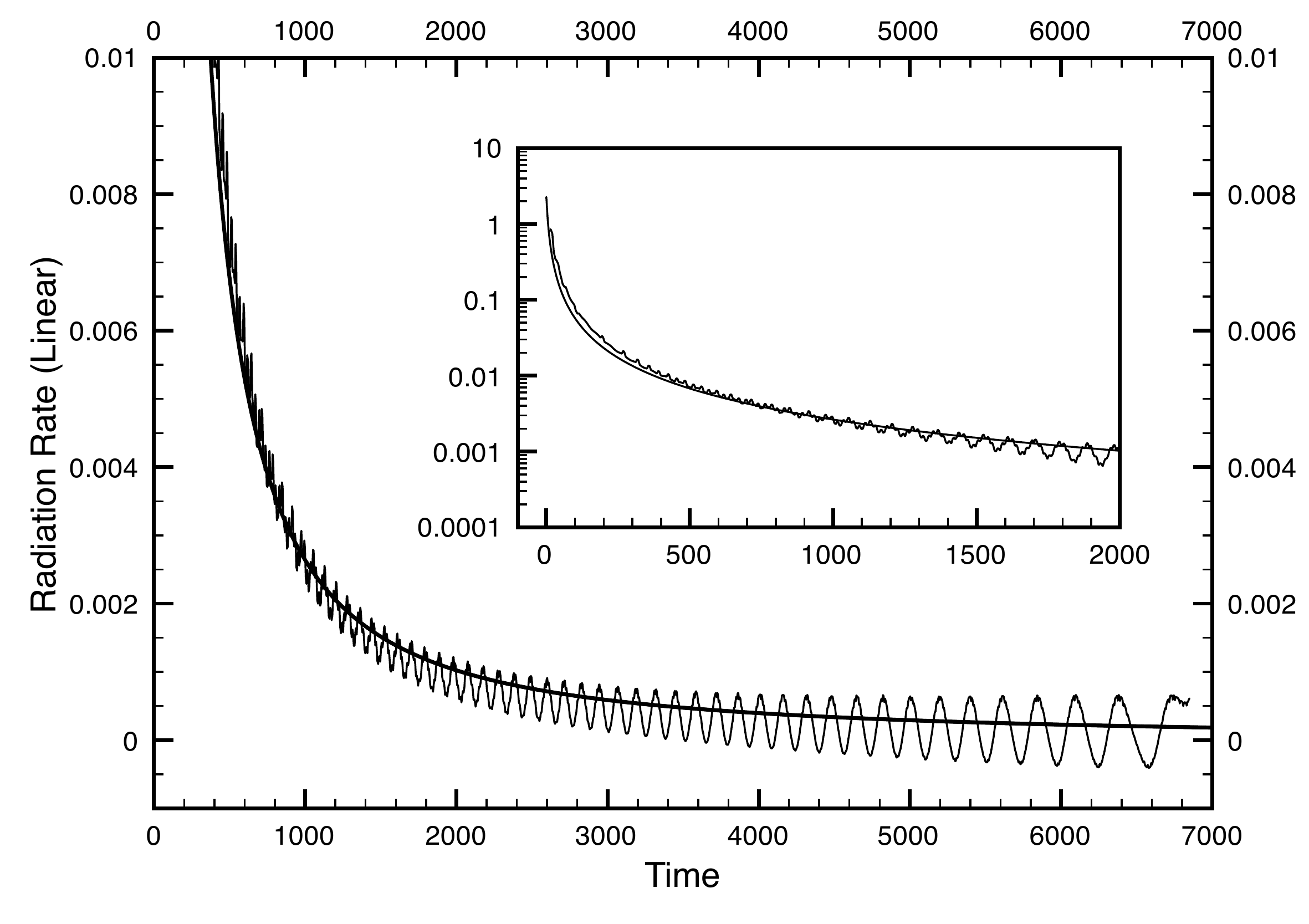}
\renewcommand{\baselinestretch}{1.0}
\caption[comp_rad_lin]
{Comparison of theoretical (continuous line) vs. numerical (wavy line) radiation
rate for an oscillon with
initial conditions ($A_0 = 2$, $R_0 = 2.86$) showing excellent agreement.  The
inset, which is plotted on a log scale, makes it clear that the theory correctly
reproduces the rapid initial drop in radiation rate over many orders of
magnitude; the linear scale on the larger graph shows that the theory correctly
reproduces the extremely small (but finite) radiation rate towards the end of
the oscillon's life [$(A_r,R_r) \simeq (.92, 3.25)$].}
\label{fig:comp_rad}
\end{figure}

\begin{figure}
\includegraphics[width=.45\textwidth,height=2.3in]{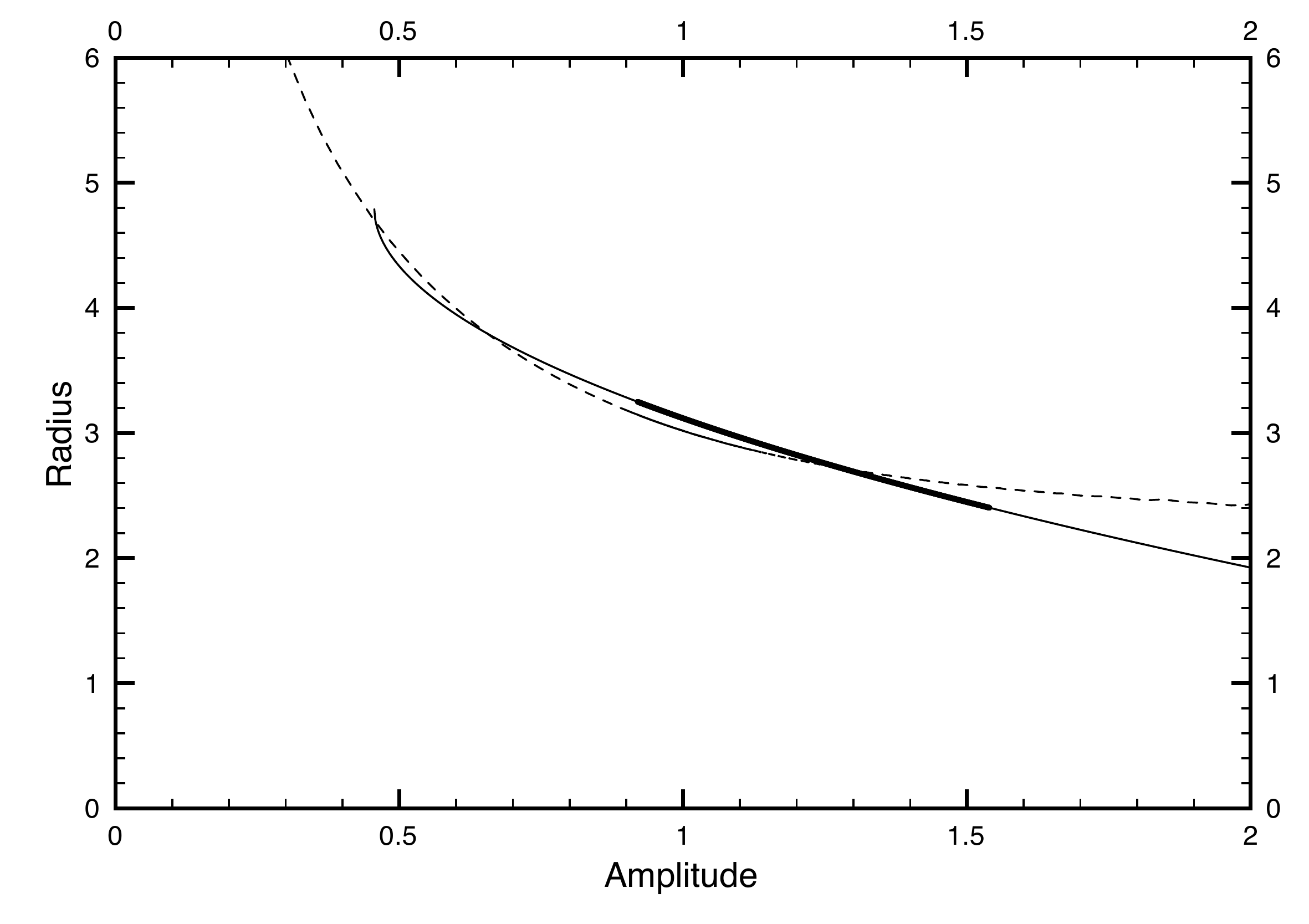}
\renewcommand{\baselinestretch}{1.0}
\caption[trajcompnum]
{Comparison of theoretical trajectory (solid curve) vs. numerical trajectory
(dashed curve) for the longest-lived oscillon ($A_0 = 2$, $R_0 = 2.86$) showing
very good agreement during the more stable phase of the oscillon's life ($A
\lesssim 1.5$).  The great increase in density of data points in the dashed line
in the range $.9 \lesssim A \lesssim 1.5$ is due to the prolonged period of time
spent in this region by the oscillon (i.e., the ``plateau'' phase).  The end
point where the dashed line again becomes dashed ($A \simeq .9$, $R \simeq 3.2$)
signals the numerical decay point of the oscillon.  The thick segment of the
solid line highlights the portion of the theoretical trajectory during the
low-radiation
plateau phase; the end of the thick segment $(A_r, R_r) \simeq (.92, 3.24)$
marks the theoretical decay point, showing very good agreement.  It is
interesting to note that, even after the oscillon decay at $A \simeq .9$, the
remaining field configuration continues to tend to the attractor point at
$(A_{\infty}, R_{\infty}) \simeq (.456, 4.79)$, as does the theoretical curve.}
\label{fig:trajcompnum}
\end{figure}

As was done in Eqs. \ref{Etlong} and \ref{Along}, our method allows us to
investigate
an oscillon evolving along a particular trajectory in detail.  As an
example, we consider the longest-lived oscillon in this model, obtained
with the initial parameters ($A_0=2;~R_0=2.86)$.  Using the information from the
curve in Fig. \ref{fig:lifeenergy}, we can find the
trajectory whose lifetime corresponds to this oscillon ($\mathcal{T}_{\rm{life}}
\simeq 7100$), marked as the thicker line in Fig. \ref{fig:trajectory} (the
coordinates of the reference (or decay) point are $A_r \simeq .92$ and $R_r
\simeq 3.25$).  We then use Eqs. \ref{rhoXfinal3} and \ref{gammaX} to calculate
$\rho_X$ and $\gamma_X$ for any parameter of interest.  Then,
combining these values with Eqs. \ref{Et} and \ref{exp2}, we can compute the
amplitude (Eq. \ref{Along}), radius, frequency, energy (Eq. \ref{Etlong}), and
radiation rate as functions of time and compare the results with the numerical
values.  The results are plotted
in Figs. \ref{fig:comp_afr}, \ref{fig:comp_nrg}, and \ref{fig:comp_rad}, showing
excellent agreement.  We can also plot the theoretical prediction for the
trajectory of this
oscillon in the $(A,R)$ plane (shown by the thicker line in Fig.
\ref{fig:trajectory}), and compare it to the numerical value, as shown in Fig.
\ref{fig:trajcompnum}.

In the next section, we will complete the characterization of this system by
deriving an expression for the frequency of the superimposed oscillation
observed in, for example, Fig. \ref{fig:spectrum_amprad}, which seems to be
connected with the oscillon decay process: the larger the amplitude of the
superimposed oscillation, the shorter the lifetime.

\section{Analysis of Oscillon Stability as a Function of Time}

Recall from the first chapter that the three conditions representing an oscillon
before decay, at the point of decay, and after decay, respectively, are:
\begin{align}
\label{death1}
\omega_{\rm{gap}} &> \Gamma_{\rm{lin}}; \\ \nonumber
\omega_{\rm{gap}} &= \Gamma_{\rm{lin}}; \\ \nonumber
\omega_{\rm{gap}} &< \Gamma_{\rm{lin}}.
\end{align}

In this section, we seek to investigate the concept of oscillon stability in
more
depth.  In doing so, we will obtain a more general formulation of Eqs.
\ref{death1} which will provide a precise measure of the oscillon's stability
when $\omega_{\rm{gap}} \ne \Gamma_{\rm{lin}}$.  We will then see that an
expression for the frequency of the superimposed oscillation seen in Fig.
\ref{fig:spectrum_amprad}, which is clearly related to stability, will naturally
emerge.

In deriving the equations governing the radiation rate
and lifetime in the previous sections, we made the simplifying assumption that
the
oscillons under study are long-lived.  Mathematically, this assumption is
employed in approximating the series expansion in Eq.
\ref{theorem} by its first term, yielding Eq. \ref{fourier3}.  The more stable
the oscillon, the smaller a given term in the series expansion will be relative
to the
term before it.

Let $\nu_n$ denote the magnitude of the $n$th term in the series of Eq.
\ref{theorem}.  In this section, instead of assuming that $\nu_n \gg \nu_{n+1}$,
we will compute the fractional difference between two adjacent terms and
take the result to be a natural measure of the stability of the oscillon.

\begin{figure}
\includegraphics[width=.45\textwidth,height=2.3in]{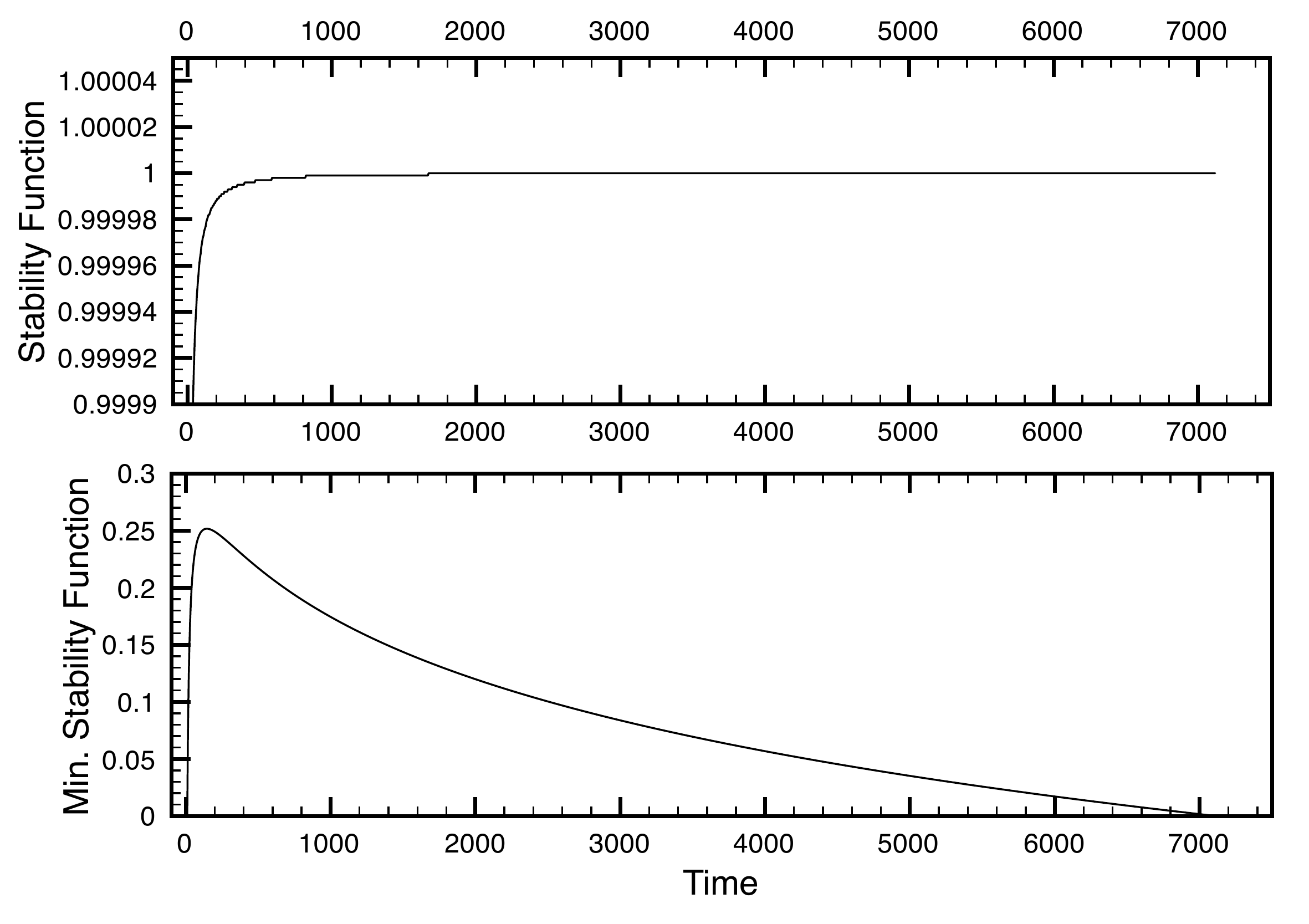}
\renewcommand{\baselinestretch}{1.0}
\caption[stabilityfunctions]
{The top graph shows the theoretical calculation of $\Sigma$ vs. time and the
bottom shows $\Sigma_{\rm{min}}$, both for the oscillon with $(A_r,R_r) \simeq
(.92, 3.25)$.  The stability measured by $\Sigma$ is clearly related to the
radiation rate of the oscillon: this kind of stability \textit{increases} in
time, since the radiation rate decreases.  On the other hand, the stability
measured by $\Sigma_{\rm{min}}$ is related to the resistance of the oscillon to
decay: this kind of stability \textit{decreases} in time as the oscillon moves
closer to the line of existence.}
\label{fig:stabilityfunctions}
\end{figure}

Define the (dimensionless) stability function $\Sigma$ from two adjacent terms
in the series of Eq.
\ref{theorem} as
\begin{equation}
\label{sigman}
\Sigma \equiv \frac{\nu_n - \nu_{n+1}}{\nu_n}.
\end{equation}
When the oscillon is highly stable, $\nu_n \gg \nu_{n+1}$, and
$\Sigma \rightarrow 1$; conversely, as $\nu_{n+1} \rightarrow \nu_{n}$
(causing the series in Eq. \ref{theorem} to fail to converge) then $\Sigma
\rightarrow 0$.
In Appendix E (see Eq. \ref{sigmanfinal}) it is shown that, for any value of
$n$,
\begin{equation}
\label{sigmar}
\Sigma = 1-\left(\frac{\Gamma_{\rm{nl}}}{\omega_{\rm{gap}}}\right)^2.
\end{equation}
Eq. \ref{sigmar} is refered to as the stability function; as would be expected,
it involves the ratio between $\omega_{\rm{gap}}$ and $\Gamma_{\rm{nl}}$.  Using
Eq. \ref{gammanl} for $\Gamma_{\rm{nl}}$, we can plot Eq. \ref{sigmar} for the
longest-lived oscillon (using Eqs. \ref{rhoXfinal3} and \ref{gammaX}).  This is
shown in the top graph of Fig. \ref{fig:stabilityfunctions}.  The extreme
closeness of $\Sigma$ to unity for most of the oscillon's life, when compared to
Eq. \ref{sigman}, verifies that we are quite justified in assuming $\nu_n \gg
\nu_{n+1}$.

The minimum stability allowed at a given time, $\Sigma_{\rm{min}}$, is attained
when $\Gamma_{\rm{nl}}$ is at its maximum, namely, when $\Gamma_{\rm nl} =
\Gamma_{\rm{lin}}$,
\begin{equation}
\label{sigmamin}
\Sigma_{\rm{min}} =
1-\left(\frac{\Gamma_{\rm{lin}}}{\omega_{\rm{gap}}}\right)^2.
\end{equation}

Now note that the conditions in Eq. \ref{death1} can be written in terms of
$\Sigma_{\rm{min}}$ as
\begin{align}
\label{death2}
\Sigma_{\rm{min}} &> 0; \\ \nonumber
\Sigma_{\rm{min}} &= 0; \\ \nonumber
\Sigma_{\rm{min}} &< 0,
\end{align}
respectively.  In the bottom graph of Fig. \ref{fig:stabilityfunctions}, we plot
$\Sigma_{\rm{min}}$ for the longest lived oscillon (again, using Eqs.
\ref{rhoXfinal3} and \ref{gammaX} and with $\Gamma_{\rm{lin}}$ given by Eq.
\ref{lineargamma}).

It is clear that one can interpret the stability $\Sigma$ as a measure of the
radiation rate of the oscillon: the radiation rate
decreases in time, so the stability \textit{increases}.  On the other hand,
$\Sigma_{\rm{min}}$ measures the resistance
of the oscillon against spontaneous decay: as the oscillon evolves and
approaches the
line of existence, this kind of stability \textit{decreases}. 

We will now write the conditions in Eqs. \ref{death1} and \ref{death2} in a
third and final form.  First, observe that we can write $\Sigma_{\rm{min}}$ as
\begin{equation}
\label{sigmaomegas}
\Sigma_{\rm{min}} = \left(\frac{\omega_{\rm{mod}}}{\omega_{\rm{gap}}}\right)^2,
\end{equation}
where
\begin{equation}
\label{omegasdef}
\omega_{\rm{mod}} \equiv \sqrt{\omega_{\rm{gap}}^2 - \Gamma_{\rm{lin}}^2}.
\end{equation}
In terms of $\omega_{\rm{mod}}$, the conditions in Eq.
\ref{death2} becomes
\begin{align}
\label{death3}
\omega_{\rm{mod}} & \in \Re,~\neq 0; \\ \nonumber
\omega_{\rm{mod}} &=0; \\ \nonumber
\omega_{\rm{mod}} & \in \Im,
\end{align}
respectively.  In other words, if we consider the quantity $B(t) \equiv
e^{i\omega_{\rm{mod}}t}$, then the real part of $B(t)$ before the decay, at the
decay point, and after the decay are
\begin{align}
\label{Bbefore}
B(t) &= \cos\left(\sqrt{\omega_{\rm{gap}}^2 - \Gamma_{\rm{lin}}^2}t\right); \\
\nonumber
B(t) &= 1; \\ \nonumber
B(t) &= e^{\pm\Gamma_{\rm{lin}}t},
\end{align}
respectively, where the last condition follows since, after the decay is
initiated, $\omega_{\rm{gap}}$ tends to zero, making $\omega_{\rm{mod}}$ tend to
$\pm i\Gamma_{\rm{lin}}$.

Therefore, we can conclude that $\omega_{\rm{mod}}$ is a special frequency
associated with the decay of the oscillon whose value decreases in time and,
at a certain point, becomes imaginary, signaling the oscillon's final demise
with
timescale on the order of the linear decay width (Eq. \ref{Bbefore}).  In fact,
as mentioned previously, such a phenomenon is commonly observed in, for example,
Fig. \ref{fig:spectrum_amprad}.  Specifically, there exists a modulation
oscillation whose frequency tends to decrease as time progresses, until, at a
certain point, the oscillon decays with
width $\sim \Gamma_{\rm{lin}}$.

\begin{figure}
\includegraphics[width=.45\textwidth,height=2.3in]{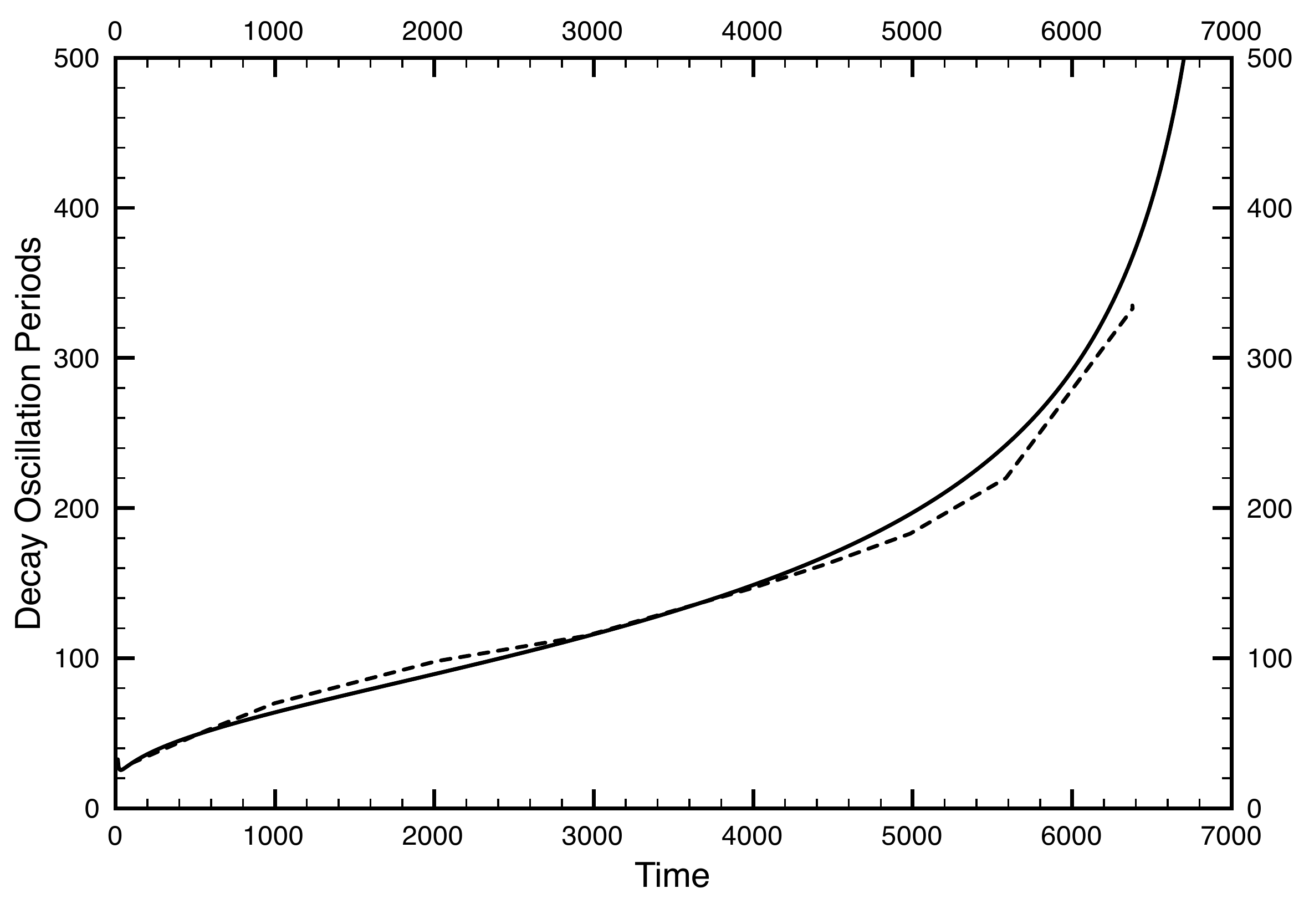}
\renewcommand{\baselinestretch}{1.0}
\caption[omega_S]
{The solid curve is the theoretical calculation of $\mathcal{T}_{\rm{decay}}$
for $(A_r,R_r) \simeq (.92, 3.25)$.  The dashed curve is the numerically
measured period of the superimposed oscillation, showing very good agreement.}
\label{fig:omega_S}
\end{figure}

In Fig. \ref{fig:omega_S} we plot the period
\begin{equation}
\label{decayomega}
\mathcal{T}_{\rm{decay}} \equiv \frac{2\pi}{\omega_{\rm{mod}}} =
\frac{2\pi}{\sqrt{\omega_{\rm{gap}}^2 - \Gamma_{\rm{lin}}^2}},
\end{equation}
along with the numerically measured period of the superimposed oscillation.  As
shown, $\mathcal{T}_{\rm{decay}}$ quite accurately reproduces this frequency.

In conclusion, we now have three separate (yet equivalent) formulations of the
condition for oscillon decay.  The first says that the nonlinear and
linear peaks must significantly overlap (Eqs. \ref{death1}).  The second says
that the measure of oscillon stability must fall to zero (Eqs.
\ref{death2}).  The last states that the modulation frequency
$\omega_{\rm{mod}}$ must become imaginary (Eqs. \ref{death3}).

\section{The Four Oscillon Timescales}

We will now review the four timescales associated with oscillons
encountered in our theory.  They are:
\begin{align}
\label{timescales}
&\mathcal{T}_{\rm{relax}} = -\frac{E-E_{\infty}}{\dot E} = \frac{1}{\gamma_{\dot
E}[E-E_{\infty}]^g} = \Gamma_{E}^{-1} \\
&\mathcal{T}_{\rm{decay}} =
\frac{2\pi}{\sqrt{\omega_{\rm{mod}}^2-\Gamma_{\rm{lin}}^2}} \\
&\mathcal{T}_{\rm{linear}} \sim \omega_{\rm{mass}} R^2 \\
&\mathcal{T}_{\rm{osc}} = \frac{2\pi}{\omega_{\rm{nl}}}.
\end{align}

The first, $\mathcal{T}_{\rm{relax}}$, is the relaxation time of the oscillon
and is typically the longest timescale present.  This is the timescale over
which the oscillon experiences significant change. It's net value expresses the
inverse rate of energy radiation, being thus largest where the oscillon radiates
the least, as can be seen from the
flatness of curves such as those in Fig. \ref{fig:comp_afr} and
\ref{fig:comp_nrg}.  This is linked to the lifetime by,
\begin{align}
\label{lifescale}
\mathcal{T}_{\rm{life}} &= \frac{1}{\gamma_{\dot
E}g}\frac{1}{[E_{\rm{D}}-E_{\infty}]^g} =
\frac{1}{g}\mathcal{T}_{\rm{relax}}\biggl \vert_{E = E_{\rm{D}}} \\ \nonumber
&\sim \mathcal{T}_{\rm{relax}}\biggl \vert_{E = E_{\rm{D}}}.
\end{align}
where we've used Eq. \ref{lifetime} and the fact that the dynamical exponents
are typically of order unity (see Fig. \ref{fig:g_calc}).  

The second timescale,
$\mathcal{T}_{\rm{decay}}$, is the period of the superimposed oscillations seen
in the oscillon as a result of its motion towards the line of existence.  The
third is the decay time of an object in the linear theory \cite{osc_glei2}.  The
fourth, and shortest timescale is the oscillation period of the oscillon.

\begin{figure}
\includegraphics[width=.45\textwidth,height=2.3in]{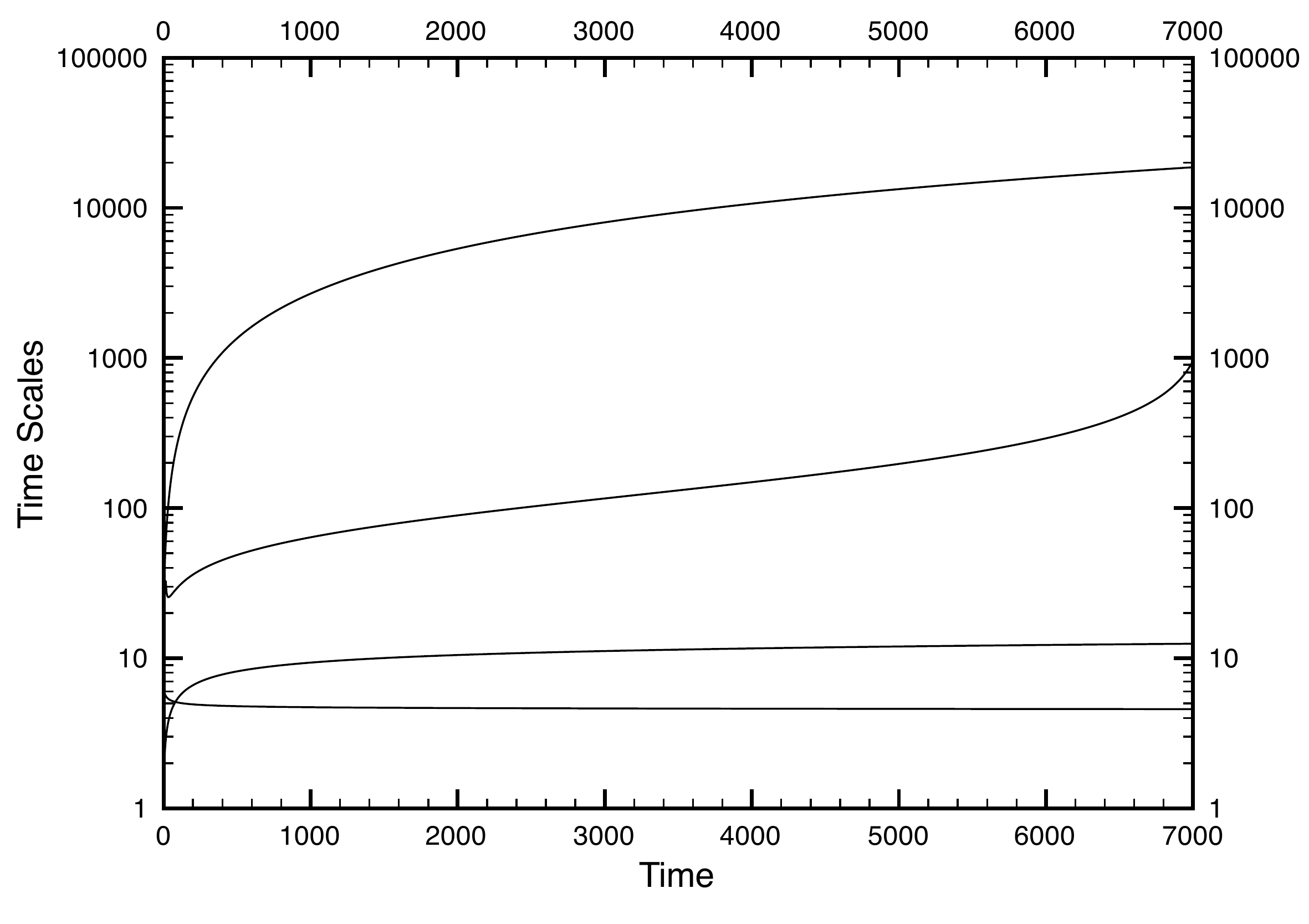}
\renewcommand{\baselinestretch}{1.0}
\caption[timescales]
{The four oscillon time scales plotted over time.  From top to bottom we have
plotted $\mathcal{T}_{\rm{relax}}$, $\mathcal{T}_{\rm{decay}}$,
$\mathcal{T}_{\rm{linear}}$, and $\mathcal{T}_{\rm{osc}}$.  It can clearly be
seen here that an oscillon is an object  governed by multiple timescales
spanning many orders of magnitude.}
\label{fig:timescales}
\end{figure}

The theoretical values of the four timescales are plotted in Fig.
\ref{fig:timescales} vs. time for the longest-lived oscillon.  Note how,
together, they span many orders of magnitude.  It is the presence of
these four widely different timescales in one single system that makes oscillons
such intriguing objects to study.

\section{Conclusions and Outlook}

In this work, we expanded upon and improved our method to compute analytically
the
decay rate and lifetime of oscillon-like configurations \cite{osc_prl}. Our
approach relies on
the comparison between the radiation spectrum of the nonlinear,
oscillon-bearing, model and its linear limit. The radiation ultimately
responsible for the oscillon decay is related to the overlap between the two
spectra: the larger the overlap, the faster the decay. We have shown that in
both $d=2$ and $d=3$ there is an attractor point in field configuration space
and that oscillons may form as configurations evolve toward that point. In
$d=3$, we may think of oscillons as unstable but very long-lived perturbations
about the attractor point. In $d=2$, oscillons may be considered as
perturbatively stable
perturbations about the attractor point. All oscillon
configurations are shown to migrate toward the attractor point as they evolve in
time.
However, in $d=3$ this point lies below the oscillon's line of
existence, and no oscillons in the models analyzed here,
double-well potentials, are absolutely stable. It remains to be seen if it's
possible to find models in $d=3$ where oscillons are perturbatively stable. In
$d=2$, the situation is
different: the attractor point lies within the oscillon region and our theory
predict that they should be at least perturbatively stable or exceedingly
long-lived. We can thus interpret the oscillons in symmetric double-well
potentials as special types of perturbations about the attractor point: those in
$d=3$ are unstable but can be
very long-lived, while those in $d=2$ are perturbatively stable. Comparison with
numerical results show that our method provides a very good quantitative
description of these configurations. We remark that there are very few examples
in the literature where analytical results provide a good description of
time-dependent, nonperturbative phenomena in relativistic field theories in
$d>1$. We have also obtained a precise criterion to establish the stability of
these configurations, as encapsulated in the oscillon stability function of Eq.
\ref{sigmamin}.

It should be born in mind that the assumptions used in this
paper, although quite general, limit the applicability of the theory.  First, it
is
known that the oscillon solution is not an exact Gaussian.  Therefore, given
that all quantitative predictions made by the theory are based on the Gaussian,
all results will possess a slight error.  The exact nature of this error is
difficult to quantify; however, it should be small given that the oscillon (in
the models studied here) is known to be well-approximated by a Gaussian. 
Another limitation of the theory lies in Eq. \ref{exp2}.  In the models studied
in this paper, the core amplitudes of oscillons with shorter lifetimes oscillate
with large amplitude about their mean value. (See, e.g., Fig.
\ref{fig:spectrum_amprad}.)
Since the energy $E(t)$ does not oscillate, it follows
that it is not possible to state, for example, that $A(t) - A_{\infty} =
\gamma_A (E-E_{\infty})^{\rho_A}$, when the oscillation amplitude of the
superimposed
oscillon (or of the radius, $R(t)$) is large.  Thus, our theory cannot make
accurate
predictions of the time-dependence of individual parameters for
oscillons which are relatively short-lived.

Given the generality of our method, we expect it to be extendable to many other
models. For example, a simple next step would be to apply it to oscillons in
asymmetric double well potentials, where their longevity is expected to increase
\cite{osc_glei1}. We currently searching for models where oscillons in $d=3$ may
have an attractor point {\it above} the line of existence. The relative location
of the attractor point with respect to the line of existence should serve as a
general criterion to determine the
longevity of time-dependent scalar field
configurations in a variety of models, including those with more than one scalar
field. Quite possibly, there may be models with two coupled fields that produce
stable oscillons even in $d=3$, at least in the sense that $\Sigma_{\rm min}
\geq 1$. Another possible extension of
the present results is the inclusion of gauge fields. As shown in Refs.
\cite{osc_U1} and \cite{osc_EW}, in both U(1) and SU(2)XU(1) models oscillons
have not been seen to decay. The question of their absolute stability remains
open and is of obvious interest. Finally, oscillons may play a key role in the
dynamics of the early universe: they may be formed during preheating after
inflation and delay thermalization; they may leave behind a
gravitational-radiation signature; and they may contribute to the dark matter
component of the cosmic energy density.
We are currently investigating how to extend the current methods to an expanding
cosmological background.

\acknowledgements
We thank Noah Graham for many useful discussions and suggestions. This work was
supported in part by a National Science Foundation grant PHY-0757124.

\appendix

\section{Linear Radiation Distribution in $d=2$}

We begin by expanding the Gaussian in eigenfunctions (zeroth-order bessel
function of the first kind) of the two-dimensional, spherically symmetric
Klein-Gordon equation:
\begin{equation}
\label{bessel1}
A(t)e^{-\frac{\rho^2}{R^2}} = \int_{0}^{\infty} b(k) J_{0}(k\rho) dk.
\end{equation}
Using the orthogonality relation
\begin{equation}
\label{bessel2}
\int_{0}^{\infty} \rho J_{0}(k\rho)J_{0}(k'\rho)d\rho =
\frac{1}{k}\delta(k-k')
\end{equation}
we can invert Eq. \ref{bessel1} to yield
\begin{equation}
\label{bessel3}
b(k) = kA(t)\int_{0}^{\infty}J_{0}(k\rho)e^{-\frac{\rho^2}{R^2}}\rho d\rho.
\end{equation}
In \cite{watson} it is shown that integration gives
\begin{equation}
\label{bessel4}
b(k) = \frac{R^2 A}{2}ke^{R^2 k^2/4}.
\end{equation}
Note that this function has the same form as the corresponding one in $d=3$. 
Hence the results in Eqs. \ref{Bomega} and \ref{omegalinear} need not be
recomputed.  To obtain $\Gamma_{\rm{lin}}$ in $d=2$ one could, in principle,
solve the equation of motion as was done in $d=3$; however, for our purposes it
will be sufficient to determine this parameter numerically (by numerically
integrating the linear Klein-Gordon equation in $d=2$ with a Gaussian initial
condition), yielding
\begin{equation}
\label{lineargammaD2}
\frac{1}{2}\Gamma_{\rm lin} = \frac{1}{\mathcal{T}_{\rm{linear}}} \simeq
\frac{.848}{\omega_{\rm{mass}}R^2} \simeq \frac{.6}{R^2}.
\end{equation}

\section{Derivation of Nonlinear Frequency Peak}

In this paper we consider spherically symmetric oscillons which can be
accurately modeled by an oscillating field configuration whose spatial profile
and amplitude of oscillation vary little over the course of an oscillation,
i.e.,
\begin{equation}
\label{separate}
\phi(r, t) \simeq A_{c}(t)P(r; R) \equiv A_{\rm{osc}}(t) A(t) P(r; R)
\end{equation}
where $\phi$ is the field,  $r$ is radial position, $t$ is time, $P(r)$ is the
spatial profile of the oscillon normalized so that $P(r) = 1$ at the origin, $R$
is a time-dependent measure of the spatial extent of the oscillon (i.e., the
``radius'') which is assumed to vary little over the period of a single
oscillation, $A_{c}(t)$ is the time-dependent oscillon core [$\phi(0,t)$],
$A(t)$ is its
time-dependent envelope of oscillation, and $A_{\rm{osc}}(t)$ is an oscillating
function which is normalized to an upper turning point of unity.

If the oscillon oscillates approximately harmonically we can write
\begin{equation}
\label{Aosc}
A_{\rm{osc}}(t) \simeq \frac{\chi}{1+\chi} + \frac{\cos(\omega_{\rm{nl}}
t)}{1+\chi}
\end{equation}
where $\omega_{\rm{nl}}$ is the time-dependent frequency of the oscillon and
$\chi$ is a
dimensionless constant which accounts for a possible non-zero center of
oscillation.

Now assume that we are interested in calculating the radiation rate of the
oscillon at $t = 0$.   Since the oscillon radiation rate depends only on the
instantaneous properties of the oscillon, writing $\omega_{\rm{nl}}(t) =
\omega_{\rm{nl}}(0)$ in Eq.
\ref{Aosc} will still yield the correct radiation rate at $t = 0$.  Combining
Eqs. \ref{separate} and \ref{Aosc} and taking the unitary cosine transform of
$A_{c}$ we have
\begin{align}
\label{fourier}
\mathcal{F}(\omega) &= \sqrt{\frac{2}{\pi}}\int_{0}^{\infty}\cos(\omega
t)A_{c}(t) dt \\ \nonumber &= \sqrt{\frac{2}{\pi}}\int_{0}^{\infty}\cos(\omega
t)\cos(\omega_{\rm{nl}}(0)) \frac{A(t)}{1+\chi} dt,
\end{align}
where in the second step we've multiplied $A(t)$ by $A_{\rm{osc}}(t)$ in Eq.
\ref{Aosc} and dropped all terms except the one proportional to
$\cos(\omega_{\rm{nl}}
t)A(t)$ which will create a finite width peak centered at the oscillon frequency
(all others are irrelevant and do not contribute to the radiation rate). We can
rewrite Eq. \ref{fourier} as
\begin{align}
\label{fourier2}
\mathcal{F}(\omega) = \frac{1}{\sqrt{2\pi}} \int_{0}^{\infty}[\cos(\omega_{+}t)
+ \cos(\omega_{-}t)] \frac{A(t)}{1+\chi} dt,
\end{align}
where $\omega_{+} \equiv \omega_{\rm{nl}} + \omega$ and $\omega_{-} \equiv
\omega_{\rm{nl}} -
\omega$.  Since each of the two cosine terms above will contribute an identical
peak (one centered at $+\omega_{\rm{nl}}$ and the other at $-\omega_{\rm{nl}}$)
and both will
contribute identically to the radiation rate, we can simply drop the one
centered
on $-\omega_{\rm{nl}}$ and multiple by two, yielding
\begin{equation}
\label{fourier3}
\mathcal{F}(\omega) = \sqrt{\frac{2}{\pi}}
\int_{0}^{\infty}\cos([\omega_{\rm{nl}} - \omega]t) \frac{A(t)}{1+\chi} dt.
\end{equation}

Now, since an oscillon's amplitude $A(t)$ tends to a finite value, denoted
$A_{\infty}$, as $t \rightarrow \infty$, we define $A_{\Delta} \equiv A(t) -
A_{\infty}$ and write
\begin{align}
\label{fourier2}
\mathcal{F}(\omega) &=
\sqrt{\frac{2}{\pi}}\int_{0}^{\infty}\cos([\omega_{\rm{nl}} - \omega]t)
\frac{(A_{\Delta} +
A_{\infty})}{1+\chi} dt  \\ \nonumber
&= \sqrt{\frac{2}{\pi}}\int_{0}^{\infty}\cos([\omega_{\rm{nl}} - \omega]t)
\frac{A_{\Delta}}{1+\chi} dt,
\end{align}
where we've dropped the transform of the $A_{\infty}$ term since it will produce
a delta function centered at $\omega_{\rm{nl}}$ which will not in anyway affect
$\mathcal{F}(\omega)$ in the region of interest (the radiation zone).

To proceed, we use the fact (found be performing successive integration by
parts)
that, for a function $f(x)$ which tends to zero as $x \rightarrow \infty$,
\begin{equation}
\label{theorem}
\int_{0}^{\infty}\cos(sx)f(x)dx  = \sum^{\infty}_{n =
1}(-1)^n\frac{f^{(2n-1)}(0)}{s^{2n}},
\end{equation}
where $f^{(m)}$ is the $m$th derivative of the function $f$.  We apply Eq.
\ref{theorem} to Eq. \ref{fourier2} and note that, for a long-lived
oscillon, the flatness of $A(t)$ implies that, for $s$ in
the tail, we need only keep the first term, yielding
\begin{equation}
\label{fourier3}
\mathcal{F}(\omega) \simeq -\sqrt{\frac{2}{\pi}}\frac{\dot
A}{1+\chi}\frac{1}{(\omega_{\rm{nl}} - \omega)^2},
\end{equation}
where we've switched the evaluations at $t = 0$ to a general time $t$, since
the above calculation can be applied to any physical time.  Eq. \ref{fourier3}
essentially states that when a peak associated with a decaying function has a
very small width then the tail goes like
$(\omega_{\rm{nl}}-\omega)^{-2}$.

It will be useful to note that $\mathcal{F}(\omega)$ in Eq.
\ref{fourier2} will be a peak whose width, denoted $\Gamma_{A}$, scales as
\begin{align}
\label{width}
\Gamma_{A} &\sim
\left(\frac{A_{\Delta}}{1+\chi}\right)^{-1}\frac{d}{dt}\left(\frac{A_{
\Delta}}{1+\chi}\right)\biggl \vert_{t = t'} =\frac{\dot A}{A -
A_{\infty}}.
\end{align}
where we've used that $\chi$ is a constant, $A_{\Delta} \equiv A-A_{\infty}$,
and $A_{\infty}$ is a constant.

\section{Overlap Function}

Consider the overlap function $\Omega(\omega)$ in Eq. \ref{overlap},
\begin{equation}
\label{overlap2}
\Omega(\omega) = \alpha \mathcal{F}(\omega) b(\omega) \equiv
\mathcal{F}(\omega)\tilde b(\omega).
\end{equation}
Since $\Omega(\omega)$ has dimension of amplitude per unit frequency (see Eq.
\ref{radamp}), as does $\mathcal{F}(\omega)$, it follows that $\tilde b(\omega)$
is dimensionless.  Therefore, one can interpret $\tilde b(\omega)$ as a
dimensionless coupling factor which modulates the distribution
$\mathcal{F}(\omega)$.  If the oscillon couples weakly to a
certain mode, $\tilde b(\omega)$ will be small at that value of $\omega$, etc.

Now, $\tilde b(\omega)$, which is proportional to the linear radiation
distribution, will be a peak whose frequency of maximum coupling
$\omega_{\rm{max}}$ is
found by
\begin{equation}
\label{omegastar}
\frac{d\tilde b}{d\omega}\biggl \vert_{\omega = \omega_{\rm{max}}} =
0.
\end{equation}

Consider now that the largest-amplitude radiation wave which can be created by a
driving force of frequency $\omega > \omega_{\rm{mass}}$ will be produced by
driving at the frequency $\omega_{\rm{max}}$.  However, the largest-amplitude
radiation wave which can be created by a given driving force will have an
amplitude on the order of the driving force itself (never larger).  Therefore,
at frequencies near $\omega_{\rm{max}}$, the amplitude of the radiation wave
$\Omega$ will be roughly equal to the amplitude of the ``driving force''
$\mathcal{F}$. In other words, $\Omega(\omega_{\rm{max}}) \simeq
\mathcal{F}(\omega_{\rm{max}})$, implying that $\tilde b(\omega_{\rm{max}})
\simeq 1$
from Eq. \ref{overlap2}.  This leads to 
\begin{equation}
\label{couple1}
\alpha \simeq \frac{1}{b(\omega_{\rm{max}})},
\end{equation}
which, when combined with Eqs. \ref{overlap} and \ref{fourier3} yields
\begin{equation}
\label{overlapgeneral}
\Omega(\omega) \simeq -\sqrt{\frac{2}{\pi}}\frac{\dot
A}{1+\chi}\frac{1}{(\omega_{\rm{nl}} - \omega)^2}
\frac{b(\omega)}{b(\omega_{\rm{max}})}.
\end{equation}

For example, in the system given by Eq. \ref{Lager1}, where $b(\omega)$ is given
by Eq. \ref{Bomega}, $\omega_{\rm{max}} = \sqrt{2+ 2/R^2}$.  This leads to
$\alpha =
e^{1/2}R/\sqrt{2}$ and, when combined with Eq. \ref{overlap}, gives
\begin{equation}
\label{couple2}
\Omega(\omega) = -\frac{R}{\sqrt{\pi}}\frac{\dot A}{1+\chi}\frac{(\omega^2
-2)^{\frac{1}{2}}}{(\omega_{\rm{nl}} -
\omega)^2}e^{\frac{1}{2}-\frac{R^2}{4}.(\omega^2-2)}.
\end{equation}
This function is plotted in the inset of Fig. \ref{fig:peaks} for typical values
of the
parameters in the system given by Eq. \ref{Lager1} in $d = 3$.  Eq.
\ref{couple2} gives the distribution (amplitude per unit frequency) of the
radiation wave emitted by the oscillon in the vicinity of a time $t$.  Its
integral with respect to frequency gives the total amplitude of the radiation
wave emitted by the oscillon, denoted $\mathcal{A}$.

Given an expression for $\Omega(\omega)$, $\delta$ and $\omega_{\rm{rad}}$ can
then be given by
\begin{align}
\label{params}
\Omega'(\omega_{\rm{rad}})&=0; \\ \nonumber
\frac{1}{2}\delta &= \left[\frac{\int
\left(\omega-\omega_{\mathrm{rad}}\right)^2\Omega(\omega)\mathrm{d}\omega}{\int
\Omega(\omega)\mathrm{d}\omega}\right]^{\frac{1}{2}},
\end{align}
where the prime denotes a derivative with respect to $\omega$.

\section{Dynamical Exponents}

The time dependence of the oscillon and all of the various parameters which
describe it is due to a \textit{single} physical mechanism, namely the emission
of radiation.  If the radiation rate is high (low), all parameters will change
rapidly (slowly).  In other words, the general nature of the time dependence of
a parameter will follow that of any other parameter.

In analogy with Eq. \ref{width}, define
\begin{equation}
\label{exp01}
\Gamma_X \propto \frac{\dot X}{X-X_{\infty}}
\end{equation}
which can be interpreted as the decay width for the parameter $X$.  Now consider
the generic parameters $X(t)$ and $Y(t)$ (which could represent amplitude,
radius, frequency, etc.).  Intuitively, we can say that the timescales
$\Gamma_{X}$ and $\Gamma_{Y}$ associated with the rates of change of the
parameters
$X(t)$ and $Y(t)$, respectively, are to be proportional,
\begin{equation}
\label{exp02}
\Gamma_{X} \propto \Gamma_{Y}
\end{equation}
for any parameters $X(t)$, $Y(t)$ (i.e., if $X(t)$ slows down by a factor of
two, then so will every other parameter). We can then write, quite generally,
$\Gamma_X \propto \Gamma_E$ for any parameter $X$, where $E$ denotes energy. 
Combining this with Eq. \ref{exp01} yields
\begin{equation}
\label{exp03}
\frac{\dot X}{X(t) - X_{\infty}} = \rho_{X} \frac{\dot E}{E(t) - E_{\infty}},
\end{equation}
where $\rho_X$ denotes the proportionality constant.  Integrating both sides of
Eq. \ref{exp03} yields
\begin{equation}
\label{exp2}
[X(t) - X_{\infty}] = \gamma_{X}[E(t)-E_{\infty}]^{\rho_{X}},
\end{equation}
where $\gamma_X$ is the constant of integration.

We will now derive an expression for the general exponent $\rho_{X}$.  First
define the new variable
\begin{equation}
\label{nuX}
\nu_{X} \equiv \ln(X - X_{\infty}),
\end{equation}
which we will henceforth employ in this Appendix, instead of $X$, to describe
the
oscillon.  When combined with Eq. \ref{exp03} this gives
\begin{equation}
\label{rhoXder}
\rho_X = \frac{d\nu_X}{d\nu_E}.
\end{equation}
During the evolution of an oscillon, the change in the coordinate $\nu_{X}$
is given by
\begin{equation}
\label{dnu}
d\nu_{X} = \nabla \nu_X \cdot d\vec \nu,
\end{equation}
where $d\vec \nu$ is a differential vector which lies tangent to the
trajectory of the oscillon in $\vec \nu$ space and $\nabla \nu_X$ is the vector
gradient of $\nu_{X}$ whose direction lies perpendicular to the ``level'' curves
associated with $\nu_{X}$.  To calculate $d\vec \nu$ we note that, by
virtue of the attractor-like nature of oscillons, the oscillon will evolve
according to a trajectory which runs perpendicular to the level curves
associated with the radiation rate.  Mathematically, we can write
\begin{equation}
\label{traj}
\nabla \nu_{\dot E} \times d\vec \nu = 0.
\end{equation}
To proceed, we will make use of our initial assumption in Eq. \ref{separate}
that
the oscillon (and hence any oscillon parameter $X$ or $\nu_{X}$) can be taken to
be a function of two degrees of freedom.  These two degrees of freedom can be
arbitrarily chosen to be any two independent oscillon parameters.  Choosing the
coordinate pair $\vec \nu = (\nu_{E}, \nu_{A})$ Eq. \ref{traj} becomes
\begin{equation}
\label{crossresult}
\frac{\partial \nu_{\dot E}}{\partial \nu_E}d\nu_A - \frac{\partial
\nu_{\dot E}}{\partial \nu_A}d\nu_E = 0
\end{equation}
Combining this with Eq. \ref{dnu} and dividing by $d\nu_E$ we have
\begin{equation}
\label{rhoXfinal1}
\frac{d\nu_X}{d\nu_E} = \frac{\frac{\partial \nu_{\dot E}}{\partial
\nu_E}\frac{\partial \nu_{X}}{\partial \nu_E} + \frac{\partial \nu_{\dot
E}}{\partial \nu_A}\frac{\partial \nu_{X}}{\partial \nu_A}}{\frac{\partial
\nu_{\dot E}}{\partial \nu_E}}.
\end{equation}

Combining Eqs. \ref{rhorad} and \ref{rhoXder} we have
\begin{align}
\label{nuEdot}
\rho_{\dot E} &= 2 - 2\rho_A + \rho_{\eta} \\ \nonumber &=
\frac{d}{d\nu_E}(2\nu_E - 2\nu_A + \nu_{\eta}) = 
\frac{d}{d\nu_E}\nu_{\dot E},
\end{align}
which implies that, up to a constant, $\nu_{\dot E} = 2\nu_E - 2\nu_A +
\nu_{\eta}$.  Substitution
of this expression for $\nu_{\dot E}$ into Eq. \ref{rhoXfinal1} and combining
with Eq. \ref{rhoXder} yields the desired result:
\begin{equation}
\label{rhoXfinal2}
\rho_X = \frac{\left(2 + \frac{\partial \nu_\eta}{\partial
\nu_E}\right)\frac{\partial \nu_{X}}{\partial \nu_E} + \left(\frac{\partial
\nu_\eta}{\partial \nu_A} - 2\right)\frac{\partial \nu_{X}}{\partial
\nu_A}}{\left(2 + \frac{\partial \nu_\eta}{\partial \nu_E}\right)}.
\end{equation}

\section{Derivation of Stability Function}

From Eq. \ref{exp2} it can be shown by differentiating that,
\begin{equation}
\label{XX}
\dot X \propto (X-X_{\infty})^{a}
\end{equation}
where $a \equiv 1+ \frac{g}{\rho_X}$.  By further differentiating, it can be
shown that
\begin{equation}
\label{GammaXX}
\frac{X^{(m)}}{X^{(n)}} \propto \left(\frac{\dot X}{X-X_{\infty}}\right)^{m-n} =
\Gamma_{X}^{m-n},
\end{equation}
where $X^{(m)}$ denotes the $m$th derivative of $X$ with respect to time.  Now,
from Eq. \ref{sigman},
\begin{equation}
\label{sigman2}
\Sigma \equiv \frac{\nu_n - \nu_{n+1}}{\nu_n} = 1-\frac{\nu_{n+1}}{\nu_n}.
\end{equation}
From Eq. \ref{theorem} we have,
\begin{equation}
\label{nuratio}
\frac{\nu_n}{\nu_{n+1}} = \frac{s^2}{f^{(2n+1)}/f^{(2n-1)}} =
\frac{(\omega-\omega_{\rm{nl}})^2}{A^{(2n+1)}/A^{(2n-1)}},
\end{equation}
where we've used Eq. \ref{fourier3} and the fact that $\chi$ is a constant. 
Combining Eq. \ref{nuratio} with Eq. \ref{GammaXX} (in the case that $X = A$) we
have
\begin{equation}
\label{nuratio2}
\frac{\nu_n}{\nu_{n+1}} \propto
\left(\frac{\omega-\omega_{\rm{nl}}}{\Gamma_{\rm{nl}}}\right)^2 =
\beta\left(\frac{\omega-\omega_{\rm{nl}}}{\Gamma_{\rm{nl}}}\right)^2,
\end{equation}
where $\beta$ is a proportionality constant and we've used that
$\Gamma_{\rm{nl}} = \Gamma_A$.  Combining Eq. \ref{nuratio2} with Eq.
\ref{sigman2} we have
\begin{equation}
\label{sigman3}
\Sigma = 1 -
\frac{1}{\beta}\left(\frac{\Gamma_{\rm{nl}}}{\omega-\omega_{\rm{nl}}}\right)^2.
\end{equation}

Eq. \ref{sigman3} is a function of $\omega$; to determine the appropriate value
of $\omega$ we note that, if $\Gamma_{\rm{nl}}$ were to attain its maximum value
of $\Gamma_{\rm{lin}}$ and if $\Gamma_{\rm{lin}} = \omega_{\rm{gap}}$, the
oscillon would decay; therefore, $\Sigma = 0$ when
$\Gamma_{\rm{nl}} = \Gamma_{\rm{lin}} = \omega_{\rm{gap}}$.  This implies that
$\beta(\omega-\omega_{\rm{nl}})^2 = \omega_{\rm{gap}}^2$.  Substitution of this
into Eq. \ref{sigman3} yields the desired result of
\begin{equation}
\label{sigmanfinal}
\Sigma = 1-\left(\frac{\Gamma_{\rm{nl}}}{\omega_{\rm{gap}}}\right)^2.
\end{equation}

 \end{document}